\documentstyle[11pt,aaspp4]{article}
\newcommand\beq{\begin{equation}}
\newcommand\eeq{\end{equation}}

\begin{document}

\title{Time-Dependent Photoionization in a Dusty Medium I: 
Code Description and General Results}

\author{Rosalba Perna\altaffilmark{1,2} and Davide Lazzati\altaffilmark{3}}

\altaffiltext{1}{Harvard Society of Fellows, 74 Mount Auburn Street, Cambridge, MA 02138}

\altaffiltext{2}{Harvard-Smithsonian Center for Astrophysics, 60 Garden Street,
Cambridge, MA 02138}

\altaffiltext{3}{Institute of Astronomy, University of Cambridge, 
Madingley Road, Cambridge CB3 0HA, UK}

\begin{abstract}
We present a time-dependent photoionization code that combines
self-consistently metal evolution and dust destruction under an
intense X-ray UV radiation field.  Firstly, we extend the mathematical
formulation of the time-dependent evolution of dust grains under an
intense radiation flux with the inclusion of the process of ion field
emission (IFE). We determine the relative importance of IFE with
respect to X-ray and UV sublimation as a function of grain size,
intensity and hardness of the incident spectrum. We then combine the
processes of dust destruction with a photoionization code that follows
the evolution of the ionization states of the metals and the relative
radiative transitions. Our code treats, self-consistently, the gradual
recycling of metals into gas as dust is sublimated away; it allows for
any initial dust grain distribution and follows its evolution in space
and time.  In this first paper, we use our code to study the
time-dependent behaviour of the X-ray and optical opacities in the
nearby environment of a Gamma-ray Burst, and show how the time
variability of the low-energy and high-energy opacities can yield
powerful clues on the characteristics of the medium in which the
bursts occur.

\end{abstract}


\section{Introduction}

The realization that Gamma-ray Bursts (GRBs) are of cosmological origin has
made them rank among the most energetic astrophysical phenomena known
to us.  Even though the total energy output is much smaller than that
of other bright sources on the sky, such as quasars, their luminosity
can be much higher.  As a result, while the surrounding region
affected by their radiation is much smaller than the corresponding
region for QSOs, the effects of the interaction can be observed on a
very short time-scale, comparable with the duration of the burst and
its longer wavelength emission.

The X-ray and UV radiation accompanying a GRB alters the equilibrium
of the medium in its close vicinity by heating and photoionizing it,
and vaporizing dust grains. The time-variability of absorption lines
and photoionization edges in GRB spectra has been discussed by Perna
\& Loeb (1998), B\"ottcher et al.  (1999) and Lazzati, Perna \&
Ghisellini (2001).  The destruction of dust by the intense UV flash
produced by the reverse shock accompanying GRBs has been treated by
Waxman \& Draine (2000) and Draine \& Hao (2002; DH in the following),
while the effect of X-rays on dust particles has been discussed by
Fruchter et al. 2001).  DH (see also Draine 2000) also computed in
detail the absorption spectrum due to the vibrationally excited levels
of the H$_2$ molecule.

The time-dependent effects resulting from the interaction of a GRB and
its longer wavelength radiation with the environment can be a powerful
diagnostics of the type of environment in which the bursts
occur. Lazzati \& Perna (2002) showed how the time-dependent X-ray
extinction is sensitive to the density profile in the close
environment of the bursts, which is different in the various
progenitor scenarios for GRBs. On the other
hand, as discussed before, GRBs, even though have a much shorter lifetime
than QSOs, they can be far more luminous.  As such, they can
allow to probe regions that are not accessible to QSO absorption
studies due to the high absorption. It has in fact been observed (Fall
\& Pei 1993) how studies of Damped Lyman $\alpha$ (DLA) absorbers done
through QSO absorption spectra are biased against observing the
densest systems. And it might very well be the case that the outer
regions probed through these studies do not constitute a fair sample
of what the properties of these high-redshift systems are. Therefore,
absorption studies with GRBs can constitute a wonderful complement to
absorption studies made with QSOs, and give us a more complete picture
of what the properties of high redshift galaxies really are. Along
these lines, it should be mentioned a study of metal column densities
and dust content made with three GRB spectra by Savaglio, Fall \&
Fiore (2002). By comparing the inferred column densities and dust
depletion amounts with those derived in DLAs from QSO absorption
studies, they inferred that the GRBs were probing denser and dustier
regions than the ones probed by QSOs.

A proper evaluation of the densities and dust content of the GRB close
environment requires a knowledge of how GRBs affect their environment,
modifying its properties, and in particular how they influence gas and
dust in their surrounding.

So far, the evolution of metals and dust under the influence of an
intense radiation field has always been treated separately. In this
paper, we present a time-dependent photoionization code that
incorporates metals and dust evolution in a self-consistent way.
Firstly, we improve the photoionization code developed by Perna,
Raymond \& Loeb (2000) and Perna \& Raymond (2000) with the addition
of a proper treatment of radiative transfer in optically thick media
and the inclusion of the opacity of the H$_2$ molecule.  We then
extend the mathematical formulation of the time-dependent evolution of
dust grains under an intense radiation flux with the inclusion of the
process of ion field emission (IFE). This process becomes important
for hard spectra of the incident radiation. We combine the processes
of dust destruction (UV and X-ray sublimation and IFE) with the
photoionization code that includes H, H$_2$ and the 12 most abundant
astrophysical elements. Our code follows, both in space and in time,
the fractions of metals that are depleted into gas and those that are
in gaseous phase.  As dust is being destroyed, metals are gradually
being recycled into gas.

Our code allows for any initial distribution of the grain sizes, and,
for each grain size and wavelength, opacities are evaluated by
interpolation on the grids of opacities computed by Draine \& Lee
(1984; DL in the following) and Laor \& Draine (1993).

While on one side showing how GRBs modify their surroundings (and
hence the inferred properties of the environment as measured at later
times), we also show how monitoring the time-variability of several
observables during the GRB event can yield powerful information on the
characteristics of the environment.  In particular, in this paper, we
discuss the simultaneous behaviour of the X-ray opacity and the
optical opacity, and show how they depend on the size and density of
the region. In two companion papers (Perna, Lazzati \& Fiore 2002, paper II,
and Lazzati \& Perna 2002, paper III) we will be discussing,
respectively, the time-dependence of the optical opacity on the
distribution of dust grain sizes in the medium and the effect of
time-dependent ionization on the appearance of X-ray spectra, 
focusing on the measure of the continuum absorption (equivalent column
density, see Lazzati \& Perna 2002) and on the appearance of enhanced
metallicities (or deeper metal edges) similarly to the case of warm
absorbers (Done et al. 1992; Zdziarski et al. 1995).

This paper is organized as follows: in \S 2 we discuss the various
dust destruction mechanisms and we derive (\S 2.2) a mathematical
expression to describe the evolution of the grain size due to the
process of IFE.  In \S 3 we describe how the code handles, during the
process of dust sublimation, the transfer of metals from dust into
gas, and how the initial dust distribution evolves in time. The
initial conditions for dust content and metal depletion pattern are
described in \S 4, while in \S 5 we discuss the photoionization and
photodissociation of H$_2$ and how the code handles radiative transfer
in optically thick regions. The evolution of metals is described in \S
6, while the basics of the radiative transfer and the dust opacities
are discussed in \S 7. Our results for the evolution of the combined
X-ray and optical opacities and their relevance for GRB environments
are presented in \S 8, while the observational perspectives (with both
current and future instrumentation) for detection of the
time-dependent effects that we describe, are discussed in \S 9.
Finally, our work is summarized in the last section, \S 10.

\section{Dust Destruction}

There are several processes that can contribute to dust 
destruction\footnote{In this work we do not include any of the mechanisms
for dust formation as the timescale for this process is much longer than
what we are interested in here (e.g. Salpeter 1977).} 
by the GRB X-ray prompt emission and (possible) UV flash.  In this
section, we analyze each of them in detail, and discuss their relative
importance depending on the spectral shape of the flux.  We consider
here the optically thin case, where simple relations can be derived,
and the dependence of the various effects on the illuminating spectrum
can be better illustrated.  Therefore, throughout this section, we
will neglect the decrease of the ionizing flux due to opacity of the
material at smaller radii. We will fully take this into account in the
numerical computation of the radiative transfer.

\subsection{Thermal sublimation}

Dust grains can be destroyed through thermal sublimation. This happens
when the grains absorb energy faster than they manage to radiate away.
The excess energy then goes into breaking the bonds that hold atoms to
the surface of the grains. The sublimation rate of a grain with
temperature $T$ and radius $a$ can be approximated by (Guhathakurta
\& Draine 1989)
\begin{equation} 
\frac{da}{dt}=-\,n^{-1/3}\,f_0 \,e^{-B/kT}\;,
\label{eq:dadt1}
\end{equation}
where $n$ is the grain atomic density\footnote{Note that we use atomic and not molecular
densities since sublimation takes place atom by atom.}, and $f_0$ and
$B$ are constants that depend on the composition of the grain.  This
process is described by Waxman \& Draine (2000), who however consider
only dust heating by UV photons.  In this case, they can write the
heating term as:
\begin{equation}
{{dE_{\rm UV}}\over{dt}} = {{L_{1-7.5}}\over{4}}{{a^2}\over{R^2}}\;,
\end{equation}
where $L_{1-7.5}$ is the UV luminosity\footnote{At energies above 7.5 keV,
photons are mainly absorbed by H$_2$ and then H (when $E\ge 13.6$ eV), 
for H-nuclei densities $n_{\rm H}\ga 10^2$ cm$^{-2}$.} 
in the range [1-7.5] eV, $a$ is
the grain radius and $R$ is the distance from the photon source. 
Whereas in the code the variation in the spectrum at a given radius due to 
the change in the opacity at smaller radii (because of the
gradual grain evaporation) is
self consistently taken into account, some useful numbers and
relations can be derived in an optically thin cloud assuming a
power-law spectrum of the heating radiation $L(\nu)=L_0 \,
\nu^{-\alpha}$. In this case, the UV heating term can be written as:
\begin{equation}
{{dE_{\rm UV}}\over{dt}} = {{L_0\,c_{\rm UV}(\alpha)}\over{4}}
{{a^2}\over{R^2}}
\end{equation}
where $c_{\rm UV}(\alpha)$ is defined as:
\begin{equation}
c_{\rm UV}(\alpha,a) \equiv {{(1.8\times10^{15})^{1-\alpha}-
(2.4\times10^{14})^{1-\alpha}}\over{1-\alpha}}\,q_{UV}(a)\;.
\end{equation}
and $q_{UV}$ takes into account the fraction of the incident flux
that is actually absorbed by the grain. To estimate this value we have
used the numerical results of Draine \& Lee (1983)
and Laor \& Draine (1993)\footnote{Tables can be found
at {\tt http://astro.Princeton.EDU/$^\sim$draine/dust/dust.html}.}.

In addition to this heating process, the grain temperature is
partially modified by the photoelectric absorption of X-ray photons.
As discussed in DH, a dust grain under heavy X-ray illumination is
initially charged up to a potential of $3\,a_{-5}(S/10^{11}\;{\rm dyn}\;
{\rm cm}^{-2})^{1/2}$ kV, 
where $a_{-5}\equiv a/(10^{-5}\,{\rm cm})$, and $S$ is the tensile strength of the material. 
After this
moment, any ionization corresponds to the emission of one or more
ions. This process is called Ion Field Emission (IFE). If the dust
grain is charged, however, not all the photoelectrons will have enough
energy to escape the grain. For this reason, the residual kinetic
energy of all the photoelectrons that cannot escape the grain is
transformed into heat.

Consider a simplified expression for the photoelectric cross
section of the $K$ electrons of an atom:
\begin{equation}
\sigma(\nu)=\sigma_{\rm th} \, \left({{\nu}\over{\nu_{\rm th}}}\right)^{-3},
\;\;\;\; \;\nu\ge\nu_{\rm th}\;.
\end{equation}
We can write the X-ray thermal heating term as:
\begin{equation}
{{dE_{\rm X}}\over{dt}} = {{n\,a^3}\over{3\,R^2}}
\int_{\nu_{\rm th}}^{\nu_{\rm th}+\nu_V}
{{L(\nu)}\over{\nu}} (\nu-\nu_{\rm th}) \, \sigma_{\rm th}
\left({{\nu}\over{\nu_{\rm th}}}\right)^{-3} \, d\nu\;,
\end{equation}
where $n$ is the density of the dust grain in cm$^{-3}$, and
$\nu_V=7.2\times10^{17}\, a_{-5}$ Hz is the frequency of photons with 3 keV
energy (the energy above which the photoelectron is free to escape and
does not give any thermal energy to the dust grain).  We need to
remark here that the above equation is not valid for $a\gg10^{-5}$ cm,
since the dust particles would then become optically thick in the soft
X-ray regime.

Using the power-law spectrum defined above, and considering more
transitions, we can write:
\begin{equation}
{{dE_{\rm X}}\over{dt}} = {{L_0\,n\,a^3}\over{3\,R^2}} \sum_i c_i(\alpha)
\end{equation}
where the sum is over the considered transitions and the single
transition coefficient $c_i(\alpha)$ is defined as:
\begin{equation}
c_i(\alpha) \equiv
\sigma_{\rm th}\,\nu_{\rm th}^3 \left\{
{{1}\over{2+\alpha}}\left[
\nu_{\rm th}^{-2-\alpha} - (\nu_{\rm th}+\nu_V)^{-2-\alpha} \right] +
{{\nu_{\rm th}}\over{3+\alpha}}\left[
(\nu_{\rm th}+\nu_V)^{-3-\alpha}-\nu_{\rm th}^{-3-\alpha} \right]
\right\}\; .
\end{equation}

Let us now specialize to Galactic-type dust, which is believed to be a
mixture of silicate and graphite grains (Mathis, Rumpl \& Nordsieck
1977). In particular, let us consider the silicate grains made of
molecules\footnote{As a matter of fact, the silicate grain is made
by a lattice of atoms rather than by bounded molecules. Here and in
the following we use the term ``molecule'' for convenience.} of
MgFeSiO$_4$ (also called olivine; Draine
\& Lee 1984), and graphite (which is Carbon).  For each of the
elements which make up these molecules, we consider only the
$K_\alpha$ photoionization of the neutral atom. In principle the grain
can be ionized also by higher orbital photoionizations. Since,
however, these have a threshold frequency which is much smaller than
the potential barrier of the grain at the IFE threshold, their
contribution to the grain ionization is negligible.  Since the inner
orbit potentials and cross sections of photoionization are only
marginally affected by the molecular bounds, we use the atomic cross
section of Verner \& Yakovlev (1995) for the elements within the
grains.

It is interesting to evaluate the relative efficiency of the two
heating terms in the idealized case of an optically thin cloud of dust
illuminated by a power-law spectrum. We have:
\begin{equation}
{{{dE_{\rm X}}\over{dt}}\over{{{dE_{\rm UV}}\over{dt}}}} =
{4\over3}\,n\,a\,{{\sum_i c_i(\alpha,a)}\over{c_{\rm UV}(\alpha)}}\;.
\end{equation}
Since the absorption cross section of Carbon is larger than that
of the silicate molecule, the relative sublimation efficiency of
graphite as compared with silicate increases as the spectrum gets
harder. Also, since X-ray heating is a volume term while UV
heating is mainly a surface term (with numerical corrections), larger
grains will be predominantly heated by the X-rays while small grains
will be heated predominantly by the UV radiation.  The result of the
above equation can be computed numerically, using the density
values $n_{\rm
Sil}=8.05\times 10^{22}{\rm cm}^{-3}$ and $n_{\rm Gra}=1.13\times
10^{23}{\rm cm}^{-3}$.  The results for both types of grains are shown
in Fig. 1.

\subsection{Ion Field Emission}

Once the dust grain is charged up to a potential
$V_{\max}=3\,a_{-5} (S/10^{11}\;{\rm dyn}\;
{\rm cm}^{-2})^{1/2}$~kV, each further ionization will lead to the
emission of one or more ions in order to avoid further increase in the
surface electric field (Muller \& Tsong 1969; Draine \& Salpeter 1979). 
This process is in competition with the
Coulomb Explosion (CE), i.e. the fragmentation of the grain into two
smaller particles (Fruchter et al. 2001).  If an initial CE takes
place, it sets in a runaway process, and the grain is rapidly and
completely dissociated into atoms, since the remnant of the first
explosion will be above threshold for a further explosion. CE is
therefore a much more effective destruction mechanism. It is also
possible that grains are originally a cluster of sub-grains of the two
different kinds, loosely bound by coesion forces. In this case, a CE
may fragment a clustered grain into its basic components that would be
consequently sublimated by IFE processes. 
The importance of CE versus IFE is uncertain. Waxman \& Draine
argue that the tensile strength of submicron grains is typically
high enough ($S > 10^{11}$ dyn cm$^{-2}$) that IFE is likely to dominate
over CE in highly charged grains. Fruchter et al. however consider that
the intense flux bombarding the grains is likely to damage the
crystalline structure of the grains themselves, therefore reducing their
tensile strength. Draine \& Hao further argue that ``while chemical
bonds will undoubtedly be disrupted by ionization, it seems likely
that chemical bonds will be promptly reestablished in the warm grain, i.e.
'annealing' will take place''.  In this case the grains would be able
to sustain IFE. Here, following Draine \& Hao, we assume that CE
does not play a relevant role, and IFE is the only mechanism
responsible for dust destruction of grains charged by high energy
photons.

The ionization rate of the grain, taking into account the potential
barrier, is:
\begin{equation}
{{dN_{\rm ion}}\over{dt}} =
{{n\,a^3}\over{3\,R^2}} \int_{\nu_{\rm th}+\nu_V}^{\infty}
{{L(\nu)}\over{h\nu}} \, \sigma_{\rm th}\, \left({{\nu}\over{\nu_{\rm th}}}
\right)^{-3} \, d\nu =
{{L_0\,n\,a^3}\over{3\,R^2\,h}}\,{{\sigma_{\rm th}\,\nu_{\rm th}^3}\over{3+\alpha}}
(\nu_{\rm th}+\nu_V)^{-3-\alpha}\;,
\label{eq:irate}
\end{equation}
where the rightmost term holds for the idealized optically thin cloud
and power-law spectra. We also assume that the energy that is
transferred to the grain by the photoelectron is negligible. As a
matter of fact, as the photoelectron travels inside the grain, it
interacts with the atoms and molecules and can lose energy mainly
through collisional ionizations. Adopting a cross section for
collisional ionization of $\sim 10^{-18}$~cm$^2$ for electrons with
several keV, we find that the energy loss is significant if
$a>10^{-5}$~cm. In practice, this can result in an increase of
the threshold frequency $\nu_{\rm{V}}$ that increases the efficiency
of X-ray heating of the larger grains. We have not included this
effect in our numerical calculations.

Let $N_i$ be the total ionization of the dust grain, which will then have
a total charge $N_i\,e$ and a surface electric field
$U=N_i\,e/a^2$, where $e$ is the proton charge.  
If we impose that the value of $U$ has to be constant,
we obtain
\begin{equation}
dN_i={{2U\,a}\over{e}}\,da
\end{equation}
i.e. a decrease of $a$ must be followed by a decrease of the total
charge of the grain $N_i$. This means that the emission of a
photoelectron will be, on average, followed by the emission of more
than one single ion\footnote{It is also possible that a single highly
ionized atom is emitted to balance the grain potential. We neglect
this possibility, and note that the difference is small, as long as the
grain has a radius $a\ge10^{-8}$~cm.}. If $N_G$ is the total number of
atoms of the grain, we have
\begin{equation}
dN_G = -dN_{\rm ion}+{{2U\,a}\over{e}}\,da\;.
\label{eq:dn1}
\end{equation}
and, assuming a uniform grain
\begin{equation}
dN_G=4\pi\,n\,a^2\,da\;.
\label{eq:dn2}
\end{equation}
Combining Equations (\ref{eq:dn1}) and (\ref{eq:dn2}) we obtain, for 
the rate of grain erosion due to IFE only
\begin{equation}
{{da}\over{dt}} = - {{1}\over{4\pi\,n\,a^2-{{2U\,a}\over{e}}}}
{{dN_{\rm ion}}\over{dt}}
\end{equation}
Note that the coefficient in the above equation is always positive,
and that with a crude approximation, one can assume that
$4\pi\,n\,a^2\gg{{2U\,a}\over{e}}$ for all the cases of
interest. This, combined with the above equation (\ref{eq:irate}),
would give an exponential decay of the radius of the particle with
time (if the dependency of $\nu_V$ on $a$ is neglected). 

\subsection{Dust destruction}

The results from \S 2.1 and \S 2.2 allow us to write the differential
equation governing the evolution of the radius of the dust grain
under the combination of UV and X-ray heating and IFE. 

It is likely that, when thermal sublimation is active, the grain will
be kept neutral since the ions, which have lower binding energy, will
be sublimated as soon as they are created. Thermal sublimation will
therefore completely quench the IFE process. The equation governing
the grain size evolution therefore reads:
\begin{equation}
{{da}\over{dt}}=\left\{\begin{array}{ll}
-{{1}\over{4\pi na^2-{{2Ua}\over{e}}}}{{dN_{\rm ion}}\over{dt}}\;\;\;\;\; & 
{\rm if}\;\;{{1}\over{4\pi na^2}}{{dN_{\rm ion}}\over{dt}} > n^{-1/3}f_0e^{-B/kT} \\
-n^{-1/3}f_0e^{-B/kT} \;\;\;\;\;& 
{\rm if}\;\;{{1}\over{4\pi na^2}}{{dN_{\rm ion}}\over{dt}} \le n^{-1/3}f_0e^{-B/kT}
\end{array}\right.\;,
\label{eq:dadt}
\end{equation}
where the parameters are $f_0=2\times 10^{15}$, and
$B/k=6.81\times10^4$ for silicates and $f_0=2\times 10^{14}$, and
$B/k=8.12\times10^4$ for graphite (Guhathakurta \& Draine 1989). 
The conditions in the right side of
Eqs.~(\ref{eq:dadt}) state that the equation in the second line should be used
when the rate of thermal sublimation of atoms is larger than the rate
of ionization and vice versa.  

The temperature of the grain must be computed self-consistently, 
with the balance equation:
\begin{equation}
{{dE_{\rm UV}}\over{dt}} + {{dE_{X}}\over{dt}} =
\langle Q\rangle_T\,4\pi\,a^2\,\sigma_B\,T^4 - 4\pi\,a^2 \, {{da}\over{dt}}
\,n\,B\;,
\label{eq:Tgrain}
\end{equation}
where $da/dt$ is defined by the second line of Eq.~(\ref{eq:dadt}) 
when X-ray/UV sublimation dominates, and
$\langle Q \rangle_T $ is the Planck-averaged absorption efficiency.
Waxman \& Draine (2000) provided an approximate analytical expression
for this quantity. This is shown, for various grain sizes and for both
silicates and graphite, in Figure 2, where it is compared
to the results derived in the detailed calculations by DL (thick lines
in the figure).  As our code follows the evolution of the grain
distribution, we need to have an accurate evaluation of the
Planck-averaged absorption efficiency as a function of the grains size
and temperature.  However, we found that the use of interpolation of
the DL data in computing $\langle Q\rangle_T$ during the simulations was
significantly increasing the running time of the simulations.
Therefore, we have developed a more accurate analytical approximation
for $\langle Q\rangle_T$, which is compared to the numerical results by
DL and to the approximation by Waxman \& Draine (2000) in Figure
2. The details of the approximations are provided in
Appendix A.

When IFE is dominant (or becomes so), we neglect the number of
ionizations that are necessary to bring the grain above
threshold. If the fluence of the ionizing continuum is large
enough to make IFE efficient, this is only a minor correction, since
the number of ionizations that a grain must undergo to reach the
threshold $N_{\rm{ion}}=Ua^2/e$ is small compared to the number of
ionizations required to destroy the grain,
$N_{\rm{ion}}\sim{}n\pi{}a^3$. The condition for IFE to be able
to sublimate a grain can be expressed as
\begin{equation}
R<\left[{{t\,L_0\,\sigma_{\rm{th}}\,\nu^3_{\rm{th}}}\over
{12\pi\,h(3+\alpha)\,(\nu_{\rm{th}}+\nu_{\rm{V}})^{3+\alpha}}}\right]^{1/2}
\label{eq:rmaxi}
\end{equation}
where $t$ is the duration of the ionizing continuum. If we consider a
graphite grain under the ionizing flux of a luminous GRB that produces
a total energy in the [1~eV--100~keV] range $E=10^{54}E_{54}$~erg with
a spectral index $\alpha=0$, the condition becomes
$R\lesssim5\,E_{54}^{1/2}\,a_{-5}^{-3/2}$~pc for IFE to destroy
completely a grain with initial radius $a$. In most cases, the grain
will be therefore sublimated by thermal effects, even though small
grains at large radii will be destroyed by IFE. For this reason, not
only the dust is sublimated, but the size distribution of dust grains
can be largely affected by the burst photons (see paper II).

We also assume that the recombination of free electrons onto
charged dust grains is negligible. This is justified by the fact that,
for any reasonable free electron density, the recombination rate is
much smaller than the ionization rate (Fruchter et al. 2001).

Figures 3 and 4 show the relative importance of the three effects
discussed above (i.e. UV heating, X-ray heating and IFE) in
determining the time evolution of the dust grain. This is shown for
various degrees of hardness of the photon spectrum and for both the
silicate and the graphite grain. As a general feature, grains at small
distances will be sublimated, while grains very far from the photon
source will be eroded by the ion field emission process. This is due
to the fact that, as already noted by Waxman \& Draine (2000), outside a certain radius the
grain temperature is not large enough to sublimate the dust particles
and IFE, which is effective also at very large radii, sets in 
(note that the figures have been produced with unlimited fluence, so that
there is no constraint on the duration of the ionizing continuum;
cfr. Eq.~\ref{eq:rmaxi}).  In addition, harder spectra make
X-ray heating and IFE more effective due to the larger ratio of X-ray
over UV photons.

\section{Transfer of dust into gas}

Let $dn_i/da(a,r,t=0)$ be the initial number density distribution of
grains of type $''i''$ within the region surrounding the source.  Let
$\Delta a(a,r,t)\equiv da/dt(a,r,t) \Delta t$ be the amount of size
reduction in the time interval $\Delta t$ due to the combined
processes of sublimation and ion-field emission of a grain of size $a$
at position $r$ within the region.  The mean number of molecules
within $\Delta a$ is given by\footnote{As a matter of fact, the
process of dust destruction occurs mostly through emission of single
atoms and radicals (e.g. MgO$^+$, SiO$^+$, FeO$^+$ for the olivine-type silicates). 
However, as it is not possible to know which
particular atoms or radicals are ejected within a particular time
interval $dt$, we simply assume that the average number of atoms of a
given species ejected during a given time is proportional to the mean
number of molecules contained within the destroyed grain volume, times
the number of atoms of the given species within each molecule (which
is basically the long-term mean value of the number of ejected atoms of each
given species).}
\beq
\Delta N_a(a,r,t)=4\pi(\rho/m)a^2\Delta a(a,r,t)\;,
\eeq 
where $\rho/m$ is the molecular density: $(\rho/m)_{\rm Gra}=n_{\rm Gra}$
and $(\rho/m)_{\rm Sil}=(n_{\rm Sil}/7)$ for the olivine-type silicates.
Hence the number density of molecules released within the time
$\Delta{}t$ at position $r$ is given by
\beq
\Delta N(r,t)=\int da \,\Delta  N_a(a,r,t)\, \frac{dn_i}{da}(a,r,t)=
4\pi \left(\frac{\rho}{m}\right) \int da\, a^2 \Delta
a(a,r,t)\,\frac{dn_i}{da}(a,r,t)\;,
\eeq
where $dn_i/da(a,r,t)$ is the number density of grains at time $t$ and
position $r$.

We divide the range of grain sizes in $''na''$ bins $a_k$ of size $\Delta a_k$
between $a=0$ and $a=a_{\rm max}$. 
At $t=0$, the initial grain distribution is
\beq
\frac{\Delta n_i}{\Delta a_k}(t=0,r)=\left\{
  \begin{array}{ll}
     f(a_k), & \hbox{for $a_{\rm min}\le a_k\le
     a_{\rm max} $} \\
   0,   & \hbox{otherwise}   \\
  \end{array}\right.\;,
\eeq
having assumed that the initial distribution is the same at all radii
before the source of radiation turns on. The evolution
of the number density distribution of grains of type $''i''$ is given
by
\beq
\frac{\Delta n_i}{\Delta a_k}(r,t) = \sum_{l=1}^{na} f(a_l) \;,
\label{eq:dndat}
\eeq
for all the grains of size $a_l(r,t)$ which satisfy $a_k\le
a_l(r,t)< a_{k+1}$. Basically, what Equation~(\ref{eq:dndat})
says, is that the number density in the $k$-th size bin 
at time $t$ is equal to the sum of the initial number densities for 
all of the size bins for which the decrease in size brings the corresponding grains 
$a_l(r,t)$ into the $k$-th bin. 

Let now $N_X$ be the number of atoms of element X in each of the
molecules that make up the dust grains.  The increase in the abundance
$A_X$ of element X at position $r$ within the time $\Delta t$ is given by
\beq
\Delta A_X(r,t) =N_X \frac{\Delta N(r,t)}{n_{\rm H}}\;,
\eeq
where $n_{\rm H}$ is the number density of H nuclei (i.e. number of protons).
The corresponding increase in the abundance $A_X$ can then be written
as
\beq
A_X(r,t+\Delta t)=A_X(r,t)+ \Delta A_X(r,t)=A_X(r,t)+N_X 
\frac{\Delta N(r,t)}{n_{\rm H}}\;.
\eeq
The new atoms are assumed to be singly-ionized. In fact, the average
charge per atom at threshold is smaller than unity:
$3U/(4\pi\,e\,n\,a)\sim5\times10^{-4}a_{-5}^{-1}$.  On the other
hand, neutral atoms will not experience any stress due to the grain
charge and, therefore, singly ionized atoms (or radicals) will be
ejected.

Defining the fractional increase in the abundance of element X,
$f_X(r,t)\equiv \Delta A_X(r,t)/A_X(r,t+\Delta t)$, the concentrations 
$CnX(j,r,t+\Delta t)$ of the
ions ``$j$'' of element X at position $r$ and time $t+\Delta t$ are
\beq
CnX(2,r,t+\Delta t)=CnX(2,r,t)[1-f_X(r,t)] +f_X(r,t)
\eeq
for the concentration of the singly-ionized element X, and
\beq
CnX(j,r,t+\Delta t)=CnX(j,r,t)[1-f_X(r,t)]
\eeq 
for the concentrations of all the other ions (i.e. $j\ne 2$).

The initial abundances of the elements are assumed to be
\beq
A_X(r,t=0)=A_\odot(X) [1-f_{\rm depl}(X)]\;,
\eeq
where $f_{\rm depl(X)}$ is the fraction of element X that is depleted
into dust.

\section{Initial Distribution of Dust Grains}

We consider a power law distributions of grains\footnote{The code can
handle any initial distribution of grain sizes. Here we adopt a power
law distribution as this is a good approximation for dust in our
Galaxy, and no much information is available on the shape of the
distribution in other galaxies.}
\beq
\frac{dn_i}{da}=A_i n_{\rm H}\,a^{-\beta}\;\;\;\;\;\;\; a_{\rm min}\le
a\le a_{\rm max}\;.
\label{eq:dnda}
\eeq
For extinction produced by the Galactic ISM, Mathis et al. (1977)
showed that a good approximation is obtained by taking the power law
index to be $\beta=3.5$ in the size range $a_{\rm min}\approx 0.005
\mu {\rm m}$, $a_{\rm max}\approx 0.25 \mu {\rm m}$, and for a mixture
of graphite and silicate grains\footnote{Whereas this is the typical
distribution for our Galaxy, it might not be the best representation
of the grain distributions for dense environments, where grain
coagulation might lead to more shallow slopes. However, as there is no
information of what might be a ``typical'' value of $\beta$ depending
on the density, we adopt here the more well-known Galactic value, and
perform a more complete study of the opacities for various grain
distributions in paper II.}.  The total mass contributed by dust by
the distribution ({\ref{eq:dnda}}) is
\beq
m_{\rm dust} = \sum_i \int_{a_{\rm min}}^{a_{\rm max}}da\frac{dn_i(a)}{da}
\frac{4}{3}\pi\rho_i a^3 =\frac{4}{3}\pi n_{\rm H}
\frac{a_{\rm max}^{4-\beta}}{(4-\beta)}
\left[1-\left(\frac{a_{\rm min}}{a_{\rm max}}\right)^{4-\beta} \right]
\sum_iA_i\rho_i\;.
\label{eq:md}
\eeq
This yields a dust-to-gas ratio $f_d$ (defined as the ratio between
the total mass in dust and the total mass in hydrogen)
\beq
f_d\equiv\frac{m_{\rm dust}}{m_{\rm Hyd}} = \frac{4\pi}{3 m_{\rm H}}
\frac{a_{\rm max}^{4-\beta}}{(4-\beta)}
\left[1-\left(\frac{a_{\rm min}}{a_{\rm max}}\right)^{4-\beta} \right]
\sum_i A_i\rho_i\;\equiv\chi(A_{\rm Sil}\rho_{\rm Sil}+A_{\rm Gra}\rho_{\rm Gra})\;.
\label{eq:fd}
\eeq
With the grain densities $\rho_{\rm Sil}\approx 3.3$ g cm$^{-3}$ and
$\rho_{\rm Gra}\approx 2.26$ g cm$^{-3}$ (Draine \& Lee 1984; Laor \&
Draine 1993) respectively for silicates and graphite, the mass ratio
between the two species of grains is given by
\begin{equation}
m_{\rm G_S}={{A_{\rm Gra}\,\rho_{\rm Gra}}
\over{A_{\rm Sil}\,\rho_{\rm Sil}}} = 0.685 \, {{A_{\rm
Gra}}\over{A_{\rm Sil}}}\;.
\label{eq:mdr}
\end{equation}
Using Equations (\ref{eq:fd}) and (\ref{eq:mdr}), we can then write
\begin{equation}
A_{\rm Gra} = {{f_d}\over{\chi\,\rho_{\rm Gra\,\left(1+{{1}\over{m_{\rm G_S}}}\right)}}}\;
\end{equation}
and
\begin{equation}
A_{\rm Sil} = 0.685 {{A_{\rm Gra}}\over{m_{G_S}}}\;.
\end{equation}
With the adopted values of $a_{\rm min}$, $a_{\rm max}$ and $\beta$,
we have $\chi=2.15\times 10^{22}$ g$^{-1}$ cm$^{0.5}$, 
$A_{\rm Gra}=9.62\times 10^{-26}$ cm$^{2.5}$ and
$A_{\rm Sil}=7.49\times 10^{-26}$ cm$^{2.5}$. 

At this point, we can compute the fraction of the elements depleted
into dust before the effects of the photoionizing field,
\begin{equation}
f_{\rm depl}({\rm C})={{A_{\rm Gra}\,\chi\,\rho_{\rm Gra}}\over{m_{\rm
Gra} A_\odot({\rm C})}} =
0.188{{A_{\rm Gra}\,\chi}\over A_\odot({\rm C})}
\end{equation}
for Carbon, that makes up graphite, and
\begin{equation}
f_{\rm depl}({\rm X})={{N_X A_{\rm Sil}\,\chi\,\rho_{\rm
Sil}}\over{m_{\rm Sil}A_\odot({\rm X})}} =
0.0192{{N_{\rm X} A_{\rm Sil}\,\chi}\over A_\odot({\rm X})}
\end{equation}
for each of the elements X that make up the silicates.
In the above equations,
$m_{\rm Gra}$ and $m_{\rm Sil}$ are the mass numbers (i.e. $m_{\rm molecule}/m_p$) 
of graphite and silicates, respectively. Therefore, we have 
$m_{\rm Gra}=12$ and $m_{\rm Sil}=172$.

As previously noted, we indicate by $n_{\rm H}$ the number density
of H nuclei; therefore, for a cloud whose
initial composition is a mixture of both atomic and molecular
Hydrogen, we take
\beq
n_{\rm H} = n({\rm H}) + 2n({\rm H}_2)\;.
\eeq

\section{H and H$_2$: Opacites and Photoionization Rates}

The code allows for any fraction of Hydrogen to be initially in its
molecular form, H$_2$.  We follow the processes of photoionization of
H$_2$, photoionization and photodissociation of H$_2^+$, and
photoionization of H.  Line absorption is included for Hydrogen but
not for H$_2$ and H$_2^+$. A detailed study of the trasmission
spectrum of H$_2$ can be found in DH. Here we remark that neglecting
the opacity due to line absorption by H$_2$ does not affect much the
propagation of the various destruction fronts, as strong line
absorption by H$_2$ occurs in a relatively narrow energy range ([11.2,
13.6]~eV, see e.g. DH). Opacities for the absorption lines of atomic H
are included.

In more detail, as the ionization front propagates,
rates and opacities for the following processes are computed:

\subsection{Photoionization of H$_2$: H$_2+\,h\nu \rightarrow$ H$_2^+ +e^-$}

We adopt the cross section for photoionization of H$_2$ derived by
Yan, Sadeghpour \& Dalgarno (1998):
\begin{eqnarray}
\sigma_{\rm H_2}(E)&=&10^{-16}{\rm cm}^2(-37.895 + 99.723x  - 87.227x^2 +
25.4x^3)\nonumber \\ 
&&{\rm for} \;\;15.4 < E < 18\; {\rm eV} 
\label{eq:sigH2}
\end{eqnarray}
\begin{eqnarray}
\sigma_{\rm H_2}(E)&=& 2\times 10^{-17}{\rm cm}^2 (0.071x^{-s} - 0.673 x^{-s-1} +
1.977 x^{-s-2} - 0.692  x^{-s-3})\nonumber \\ 
&&{\rm for}\;\; 18 < E < 85\; {\rm eV} 
\end{eqnarray}
\begin{eqnarray}
\sigma_{\rm H_2}(E)&=&45.57\times 10^{-24}{\rm cm}^2
(1. - 2.003/x^{0.5} - 4.806/x + 50.577/x^1.5 - 171.044/x^2\nonumber \\  
&& + 231.608/x^{2.5} - 81.885/x^3) / (E/10^3)^{3.5} 
\;\;\;\;\;\; {\rm for}\;\;  E > 85\; {\rm eV}\;,
\end{eqnarray}
where $x\equiv E({\rm eV})/15.4$ and $s=0.252$.

\subsection{Photodissociation of H$_2^+$: H$_2^+ + h\nu \rightarrow$
H$^+ +\, $H}

We adopt the fit derived by DH to the photodissociation cross section
found by von Bush \& Dunn (1972) after averaging over the H$_2^+$
vibrational distribution:
\beq
\sigma(E)=2.7\times 10^{-16} {\rm cm}^2 \left(\frac{E}{29\,{\rm eV}}\right)^2
\left(1-\frac{E}{29\, {\rm eV}}\right)^6\;\;\;\;\;\; {\rm for} E< 29\; {\rm eV}\;.
\eeq

\subsection{Photoionization of H$_2^+$: H$_2^+ +\, h\nu \rightarrow$
2H$^+ + e^-$}

We use the photoionization cross section calculated by Bates \& Opik
(1968)
\beq
\sigma(E) = 9.3\times10^{-19} {\rm cm}^2\left(\frac{E}{15.4\;
{\rm eV}}\right)^{-2}\;\;\;\;\; {\rm for} E>15.4\; {\rm eV}\;.
\eeq

\subsection{Photoionization of H: H $+\, h\nu\rightarrow$ H$^+ + e^-$}

Finally, for photoionization of atomic hydrogen, we use the cross
section (Osterbrock 1989)
\beq
\sigma(E) = 6.3\times 10^{-18}{\rm cm}^2 \left[1.34\left(\frac{E}{13.6\,
{\rm eV}}\right)^{-2.99} -0.34 \left(\frac{E}{13.6\,
{\rm eV}}\right)^{-3.99}\right]\;\;\;\;\; {\rm for} E>13.6\; {\rm eV}\;.
\eeq

\subsection{Photoionization rates}

Due to computational limitations, the radial grid over which the
radiative transfer is made will often have to be optically thick to
the processes described above when the region of study is very dense.
To deal with this problem, we adopt the numerical procedure developed
by DH; i.e. we choose the shell tickness $\Delta r$ such that an
individual shell is optically thin in dust and in the elements heavier
than H, but it might be opaque to photoionization of H$_2$ and H, and
to photodissociation and photoionization of H$_2^+$.  Rates for these
processes are computed using the modified expressions
\beq
\eta_X = \frac{\Delta r}{\Delta V}\int_{\nu_X}^\infty
\frac{L_\nu}{h\nu}\;e^{-\tau_{\nu}(r,t)}\;\frac{1-e^{-\Delta\tau_{\nu}(r,t)}}
{\Delta\tau_{\nu}(r,t)}\;
\sigma_X(\nu)\; d\nu
\label{eq:rates}
\eeq
where $\Delta V\equiv (4\pi/3)[(r+\Delta r)^3-r^3]$ is the shell
volume, $\Delta \tau_\nu(r,t)$ is the optical depth arising in the shell 
from $r$ to $r+\Delta r$ (see Eq.~{\ref{eq:tau}}), and $\tau_\nu(r,t)$ is the
optical depth up to radius $r$ (see Eq.~\ref{eq:taut}).

\section{Evolution of Metals}

Besides Hydrogen, our code includes the 12 most abundant astrophysical
elements, i.e.  He, C, N, O, Ne, Mg, Si, S, Ar, Ca, Fe, Ni. Their
abundances are taken from Anders \& Grevesse (1989).  We denote the
local number densities of the ions of the various elements by
$n^X_j(r,t)$, where the superscript $X$ characterizes the element and
the subscript $j$ identifies the ionization state.  At $t=0$, the
ionization fractions of the elements are assumed to have their
equilibrium value for an initial temperature\footnote{The evolution of
the plasma is rather insensitive to the precise value of $T_{\rm in}$
given that when the plasma is illuminated by the source it reaches a
temperature $T\gg T_{\rm in}$.}  $T_{\rm in}\sim 10^3$ K.  When the
source turns on, the abundances of the ions of the various elements are
updated by solving the system of equations
\begin{equation}
\frac{dn^X_j(r,t)}{dt}=s_{j-2}n^X_{j-2}+ q_{j-1}n^X_{j-1}
+ c_{j-1}n_{j-1}^X n_e -(q_j+c_j n_e +\alpha_j n_e)n_j^X 
+\alpha_{j+1}n_{j+1}^X n_e\;.
\label{eq:dndt}
\end{equation}
The $q_j$ and $c_j$ are respectively the photoionization and
collisional ionization coefficients of ion $j$, while $\alpha_j$ is
the recombination coefficient. Note that $s_{j-2}$ refers to inner
shell photoionization followed by Auger ionization.  The collisional
ionization rates are calculated according to Younger (1981). We
compute the terms due to photoionization by integrating
$F_\nu\sigma_\nu$ numerically.  The recombination rates are given by
the sum of the radiative and dielectronic recombination rates.  The
radiative recombination process is the inverse of photoionization, so
the rates to the ground states are computed from the photoionization
cross section with the help of the detailed balance
relation. Hydrogenic rates are used for radiative recombination to
excited levels.  The dielectronic recombination rates are taken from
Burgess (1965) with modifications to take more recent calculations
into account. Most important is the reduction due to autoionization to
excited states (Jacobs et al. 1977), with an appropriate treatment of
the weakening of this effect at higher $Z$ (Smith et al. 1985).  Since
we are dealing with a non-equilibrium situation, the ionization states
of the elements are calculated within the program itself and updated
at each position of the radial grid at each time step.
Photoionization heating and radiative cooling are calculated within
the same code, and used to update the temperature of the plasma as a
function of position and time.

\section{Radiative Transfer and Optical Depth}

We consider a region surrounding the GRB site of size $R$ and medium
density $n_{\rm H}$, and we split it up into a radial grid with steps
$\Delta{}r$.  In propagating from a point at position $r$ to another
point at position $r+\Delta r$, the afterglow flux is reduced
according to
\begin{equation}
F_\nu(r+\Delta r,t+\Delta t) = F_\nu(r,t)\exp [-\Delta \tau_\nu(r,t)]
\frac{r^2}{(r+\Delta r)^2}\;,
\label{eq:flux}
\end{equation}
where $F_{\nu}$ is in units of ${\rm {erg~cm^{-2}~s^{-1}~Hz^{-1}}}$
and $\Delta \tau_\nu(r,t)$ is the total optical depth due to the shell
at position $r$, of width $\Delta r$, and at time $t$.  This opacity,
at a given frequency $\nu$, is produced by the contribution of various
processes: absorption by H, H$_2$ and H$_2^+$, absorptions by metals
not in dust grains, and absorption by dust, that is:
\begin{equation}
\Delta \tau_\nu(r,t)= \delta\tau_{\nu,{\rm H}_2}(r,t) +
\delta\tau_{\nu,{\rm H}_2^+}(r,t) +
\delta\tau_{\nu,{\rm H}}(r,t) + \sum_{X,j}\delta\tau_{\nu,X,j}(r,t) +
\delta\tau_{\nu,{\rm dust}}(r,t)\;,
\label{eq:tau}
\end{equation}
where $\delta\tau_{\nu,X,j}(r,t)$ is the total optical depth (which
includes photoabsorption and line absorption) of the ion $j$ of
element $X$ (for H, H$_2$ and the other 12 elements included in our
code).  The photoionization cross sections are taken from Reilman \&
Manson (1979).

The contribution to the opacity by dust within a shell of width
$\Delta r$ at radius $r$ and time $t$ is given by 
\beq
\delta\tau_{\nu,{\rm dust}}(r,t) = \Delta r\,\sum_{i=1}^2\int da\;\pi a^2(r,t)
\frac{dn_i(a,r,t)}{da}[Q_{\rm abs,i}(a,\nu)+(1-g)Q_{\rm
sca,i}(a,\nu)]\;.
\label{eq:td}
\eeq
The indices $i=1,2$ in the above equation indicate the two types of
dust grains that we are considering here, i.e. silicate and graphite.
The terms $Q_{\rm abs,i}(a,\nu)$ and $Q_{\rm sca,i}(a,\nu)$ are the
efficiency factors for absorption and scattering, while $g$ is the
scattering asymmetry factor (see Laor \& Draine 1993 for more details). 
All these quantities depend on both the
energy of the incident radiation and the size of the grains.  A
detailed computation has been performed by Draine \& Lee (1983) and
Laor \& Draine (1993), based on both theoretical calculations and on
the results of laboratory experiments.  We used their tables (see
above for access information) to build grids in $\{a,\nu\}$, over
which a bilinear interpolation is performed to determine the various
coefficients for each type of grain at the needed energies and sizes.
The total optical depth up to the shell at radius $r$ is given by
\beq
\tau_\nu(r,t) = \sum_{k} \Delta\tau_\nu(r_k,t)\, ,\;\;\;\;\;\;\;\;{\rm for}
\;\;\;\;\;r_k+\Delta r\le r\; .
\label{eq:taut}
\eeq

With the inclusion of the optical depth in the trasmitted flux, the
temperature of the grains in the middle of the radial shell
$(r,r+\Delta r)$ is determined by numerically solving
Eq. (\ref{eq:Tgrain}) with the substitution $L_\nu\rightarrow
L_\nu\exp[-\tau_\nu(r,t)-0.5\Delta\tau_\nu(r+\Delta r,t)]$.

\section{Results}

In this section we show results of our numerical simulations for a
standard galactic dust law (that is with $\beta =3.5$ as power of the
initial grain distribution) and all Hydrogen in its atomic phase.
The behaviour of the opacities for various initial fractions of
Hydrogen in molecular phase and for different grain distributions will
be discussed in a following companion paper (paper II).

\subsection{Destruction fronts}

Figure 5 shows the photoionization front of Hydrogen and
the distruction fronts of both the silicate and the graphite grains.
The dust-destruction fronts are determined by finding the radius at
which there has been complete sublimation of the largest grains in the
distribution. In all cases, notice that the destruction front of
graphite lags behind that of silicate. This is because graphite is
more difficult to sublimate (see the difference in the $f_0$ value in
Sect. 2.3).
 
The three panels in Figure 5 show the evolution in different
environments, which are all characterized by the same initial column
density ($N_{\rm H}(0)=10^{22} {\rm cm^{-2}}$) and extinction
($A_V(0)= 4.5$ mag), but by different densities and sizes of the
absorbing region.  The denser the region is, the more the destruction
of dust is efficient with respect to photoionization.  This is
because, for the range of densities under consideration, the regions
through which the radiation front passes are very optically thick to
photoionization of Hydrogen, but not much optically thick to
dust. Therefore, as the density increases, the radiation front of
Hydrogen is slowed down more than that of dust destruction.

In the bottom panel of Figure 5, the absorbing region is a
dense and compact molecular cloud, with particle density $n_{\rm H}=10^4$
cm$^{-2}$ and size $10^{18}$ cm$^{-2}$. Here, for the assumed
illuminating spectrum $L_\nu=L_0\, \nu^{-0.5}$ (with $L_0$ normalized so
that the [1eV - 100 keV] luminosity is $10^{50}$ erg/s), we find that
the destruction fronts of both the silicates and graphite proceed well
ahead of the Hydrogen photoionizing front. The separation between the
dust and the Hydrogen fronts is reduced in the case of a larger
($R=10^{19}$ cm) but less dense ($n_{\rm H}=10^3$ cm$^{-2}$) region, which is
shown in the middle panel of Figure~5. This trend continues
at lower density, as it can be seen in the top panel of Figure
5, where the photoionizing front has been propagated
through an even larger ($R =10^{20}$ cm) but less dense ($n_{\rm H}=100$
cm$^{-2}$) region.

\subsection{Metal recycling from dust into gas}

As explained in \S 3, our code follows both in space and in time the
fraction of dust-depleted metals that is in dust and in gas. As the
destruction fronts propagate, dust-depleted metals are being recycled
into gas.  The bottom panel of Figure 6 shows the radial
dependence of the abundance of Fe (being released by the destruction
of the silicate grains) in gaseous phase at three different times
during the passage of the ionizing flux.  In the top panel of the same
figure, the abundance of C (which is released during the destruction
of the graphite grains) is shown at the same times.  In both cases,
the absorbing region considered is that correponding to the middle
panel of Figure 5, that is a region with density $n_{\rm H}=10^3$
cm$^{-3}$ and radius $R=10^{19}$ cm.  For this same situation, Figure
7 shows the total columns densities of O, Fe and C in
gaseous phase as a function of time. Most of the dust mass at a given
radius is recycled before the dust destruction front reaches that
radius.  This is because the smallest dust grains are sublimated away
on a much shorter time scale than the largest grains (whose complete
sublimation defines the position of the dust destruction front), and
for the steep galactic grain distribution ($\propto a^{-3.5}$) that we
have adopted here, a good fraction of the dust mass is indeed in the
small grains.

From the point of view of observations, the
recycling of metals from dust grains into gas can produce
observable consequences. For example, the grains are optically thick
in the UV regime, and therefore the ionization of the high level
electrons of metals in dust grains will be slower than for metals in
the gas phase. As a consequence, absorption lines from intermediate
ionization states of Mg, Fe and Si are observable for a longer time in
a dusty environment than in a pure gas. In principle, one may even be
able to disentangle small dust grains from big grains due to the fact
that the big grains, being thicker, are more efficient in screening
the entrained metals from the radiation field. A good candidate line
for this test is the Si IV doublet at 1394~\AA and 1402~\AA, which has
an ionization potential of $\sim 50$~eV. At larger
energies, that are relevant to X-ray observations, the grains are
usually optically thin to radiation (if extremely large grains are not
considered). In this case, distinguishing whether a metal is in dust
or in gas is observationally more challenging, especially given the
lack of detailed calculations of the photoabsorption cross sections
of atoms bound in molecules (e.g. Paerels et al 2001). 

\subsection{X-ray vs. optical extinction}

Optical and X-ray extinction are the two main observable quantities
which, once determined through spectral fitting to the data, can be
used to set constraints on the type of environment surrounding the
source. The optical extinction is typically quantified in terms of
$A_V$, the extinction in the V band, while the X-ray extinction is
quantified through the effective Hydrogen column density, $N_{\rm H}$,
usually measured under the assumption of solar abundances and cold
(i.e. neutral) material.  Even though the ratio between the effective
Hydrogen column density inferred from optical extinction, $N_{\rm H,opt}$ 
and that inferred from X-ray extinction, $N_{\rm H, X}$, cannot be
used to infer metallicities in the absorbing region (as both scale
with metallicity roughly in the same way), however, they can be used
to calibrate the dust-to-gas ratio, $f_d$ (defined in
Eq.~\ref{eq:fd}). In fact, whereas optical extinction is very
sensitive on whether metals are in gaseous phase or depleted into
dust, X-ray absorption is almost completely insensitive to
it\footnote{There are small variations in energy of the positions of
the absorption edges if the metals are bound into molecules (as in
dust grains) rather than being in gaseous phase. However these affect
very little the overall amount of absorption.}.

In this section, we study the time-dependent evolution of both the
X-ray and the optical extinction under the influence of a strong
radiation field and for different types of environments (i.e. various
densities, sizes of the absorbing regions). It should be remarked that 
the assumption here is that most of the observed absorption, both in X-ray and in optical,
is due to the immediate circumburst environment. Clearly, if our line of
sight to the source passes through a dense cloud with density much higher
than the one in the burst vicinity, then the time variable effects
in the opacities that we will describe in the following will not
be detectable, or they will be small corrections to the overall extinction.

Following Lazzati \& Perna (2002) (but see
paper III for a more detailed treatment of X-ray continuum opacity),
we parameterize the time-dependent X-ray extinction through the
effective Hydrogen column density
\begin{equation}
N_H(t) = N_H(0) \left\langle {{\tau(\nu,t)}\over{\tau(\nu,0)}} 
\right\rangle_{\rm[1-10]}\;,
\label{eq:nh}
\end{equation}
where the symbol $\langle \rangle_{\rm[1-10]}$ represents the average
over the frequency range [1--10]~keV.  The optical extinction, $A_V$,
is computed as
\beq
A_V(t) = -\log\left(\frac{F_V(t)}{F_V^0(t)}\right)\;,
\label{eq:av}
\eeq
where $F_V(t)$ is the outgoing flux in the V band at time $t$, while
$F_V^0(t)$ is the flux at the same time that would be observed if
there were no extinction.

Figure 8 shows the time evolution of the X-ray and
optical absorption (computed as described above) for the types of
environments considered in Figure 5. In all three cases the
initial column density of cold gas and of dust is the same, but its
distribution is different.  The general trend for both the X-ray and
optical extinction is that of a faster decline for a more compact,
denser region.  This effect had been originally observed by Perna \&
Loeb (1998) in their study of the time variability of the opacity of a
single line. It is simply due to the fact that, for a given source
luminosity, the closer the material is to the source, the more easily
it is ionized due to the more intense flux to which it is exposed.

Whereas the general trend in the time variability with density and
compactness of the absorbing region is the same for both X-ray and
optical extinction, the extent to which they vary depending on the
characteristics of the region is different, and this bears important
consequences for studies of the GRB environments that are based on a
comparison of the X-ray and optical extinction (e.g. Galama \& Wijers
2000).  In a small and compact region (top panel of the figure), dust
destruction is more efficient than photoionization and, as a result,
an observer that were to measure $N_H$ and $A_V$ from afterglow
fitting at later times\footnote{By ``later times'' here we mean the
times after which the most intense X-ray UV radiation from the burst
has passed through the region (on the order of several tens to several
hundreds of seconds). The later X-ray UV afterglow radiation would not
further affect the opacities much. It is however critical, especially
if the UV range is concerned, the assumption that an optical-UV flash
is present in all GRBs. Should this be false, the fluence of the
afterglow in the UV would be larger than that of the burst, and the
late afterglow radiation may still be important to modify the
ionization stage of the ISM.}  would infer a dust-to gas ratio which
is lower than the real value prior to the burst occurrence.  On the
other hand, as the density becomes lower, photoionization starts to
become gradually more efficient with respect to dust destruction, and
in this case an observer taking measurements at later times would
infer a dust-to-gas ratio that is larger than the real value. This
effect is explicitly shown in Figure 9, where we plot
the ratio between the optical and the X-ray opacity for the same types
of environments and illuminating spectrum considered in Figure
8.  The general trend is that of $A_V/N_{\rm H} <
A_V(0)/N_{\rm H}(0)$ (where $A_V(0)/N_{\rm H}(0)$ represents the value
that would have been observed had not the source altered the status of
the medium) for a short-duration burst and high densities, while
$A_V/N_{\rm H} > A_V(0)/N_{\rm H}(0)$ for a source of longer duration and lower
circumburst densities. Note that the rapid increase in the ratio
$A_V/N_{\rm H}$ at $t\sim 100$ sec for the case $n_{\rm H}=10^3$
cm$^{-3}$, $R=10^{19}$ cm is due to the fact that the Hydrogen
ionization front has reached the boundary of the region while the dust
front is still lagging behind (see Fig.5). It should
however be emphasized that in this case a prominent iron edge should
be detected in the X-ray spectrum, since the iron ionization front is
inevitably lagging behind the dust destruction one.

The trend for photoionization to become less efficient with respect
to dust destruction as the density increases, 
is further illustrated in Figure 10, where the
initial density of cold material is $N_{\rm H}=5\times 10^{23}$
cm$^{-2}$, distributed over a region of radius $R=10^{19}$
cm. Comparing the evolution of the destruction fronts and the
extinctions with the middle panel of Figures 5 and
8 (which are calculated for a region of the same size but
with density 50 times smaller), it can be seen that the difference between the
amount of extinction in X-ray and in optical is further enhanced.

In all the simulations presented so far, we have assumed an incident
power-law spectrum $L_\nu\propto \nu^{-\alpha}$, with $\alpha=0.5$,
while examining the behaviour of the low and high-energy opacities in
various types of environments.  In Figure 11, we
explore the time variability of the opacities under the effects of a
softer incident continuum, with $\alpha=1$.  A softer spectrum
accelerates the destruction of dust more than the photoionization of
the gas. This is due to the fact that, for $\alpha\ga 0.5$, the
dominant dust destruction mechanism is UV thermal sublimation (see
Fig. 2 and Fig. 3), which is essentially
dominated by the photons in the [1-7.5] eV range.  The variability of
the soft X-ray absorption $N_{\rm H}$, on the other hand, is tied to
the photoionization of Hydrogen\footnote{This is due to the fact that
the second ionization potential (i.e. the ionization potential of the
singly ionized atom) of metals that mostly contribute to the [1-10]
keV absorption, such as Fe, is close and above 13.6 eV, and therefore,
unless the absorbing region is very optically thin, the
photoionization of these heavier elements follows that of Hydrogen.},
which depends on the number of photons above 13.6 eV. Therefore, the
softer the spectrum (hence the larger the ratio between the photons in
the [1-7.5] eV range and those above 13.6 eV), the easier it is
to destroy dust with respect to Hydrogen.

The ratios between the optical and X-ray opacities, for the same types
of environments and illuminating spectrum considered in
Figure 11, are shown in Figure 12.
Again, the general trend is that of $A_V/N_{\rm H} < A_V(0)/N_{\rm H}(0)$ for a
short-duration burst and high densities, while $A_V/N_{\rm H} > A_V(0)/N_{\rm H}(0)$ for
a source of longer duration and lower circumburst densities. However, with
respect to the case $\alpha=0.5$, the switch from $A_V/N_{\rm H} < A_V(0)/N_{\rm H}(0)$ to
$A_V/N_{\rm H} > A_V(0)/N_{\rm H}(0)$ occurs for lower circumburst densities.
This is due to the increase in the efficiency of dust destruction with
respect to H photoionization for softer spectra, as explained above.

\section{Observational Perspectives}

An important issue is whether the time-dependent effects in the
opacity predicted and discussed in this paper and the two accompanying
papers are detectable with present or future instrumentation. Lazzati
\& Perna (2002) have shown that the effect of the evaporation of the
soft X-ray absorption is observable with current instrumentation in
GRBs if they explode in overdense regions within molecular
clouds. Indeed, albeit not extremely statistically compelling, the
effect is possibly observed in at least two GRB lightcurves. In the
near future, with {\it BeppoSAX} turned off and the soft X-ray camera
on board HETE-2 damaged, it is hardly possible that any better data
will be acquired. The situation is however likely to change with the
launch of the {\it Swift} satellite, forseen in 2003. This satellite,
in fact, will have for the first time an X-ray telescope, equipped
with CCD detectors, capable of repointing the location of a GRB in a
timescale of several dozens of seconds. In few cases, this repointing
should be made in even less time. Thanks to the high spectral and
temporal resolution of the X-ray telescope-CCD system, high quality
data will be available to compare with the models and confirm (or
invalidate) the preliminar results of LP02. The capabilities of this
instrument, moreover, should allow not only to follow the broad band
X-ray absorption, but also to track any individual absorption feature
present in the spectrum (Amati et al. 2000; Lazzati et al. 2001;
Lazzati, Perna \& Ghisellini 2001).

As for optical observations, the perspectives are even more
optimistic. Optical and NIR observations of afterglows are performed
now regularly, with a reaction time that is decreasing with time. Even
though very accurate optical light curves are available for a number
of GRBs, in the NIR the data are still sparse and the existence of a
dust-to-gas ratio problem is still under debate (Galama \& Wijers
2000; Stratta et al. 2002). In addition to that, it is clear from the
discussion above that the effects we describe in this paper are
evident in the early stages of the GRB history, where data in the
optical and NIR range are still unavailable. Again the situation is
going to change within two years, when {\it Swift} will be
launched. The optical telescope on board {\it Swift} will in fact
provide optical and UV photometry and spectra since the early stages
of the afterglow, if not during the GRB itself (the timescale for
repointing the optical telescope is the same as for the X-ray
one). In addition, robotic NIR telescopes from the ground are planned
(see e.g. the REM telescope, Zerbi et al. 2002). These telescopes will
provide NIR photometry on the same short timescale as the optical one,
allowing for studies on the evolution of the dust opacity and
reddening.

In conclusion, the quality of data that will be available in the near
future seems to be perfectly suited for the comparison of the
observations with the prediction of our models, allowing to understand
whether GRBs truly explode within molecular clouds and/or dense
regions, and allowing a study of the properties of these regions
in high redshift galaxies.

\section{Summary}

The discovery of the bright GRB longer-wavelength counterparts, in
particular in X-ray and UV, has brought up the problem of simultaneous
photoionization and dust destruction under an intense radiation
field. We have tackled the problem in this paper, and the
principal results of our work are the following:

\begin{itemize}

\item We have generalized the mathematical formulation of the dust destruction
mechanisms under an intense radiation field to include the process of ion field
emission, and we have discussed its relative importance with
respect to X-ray and UV sublimation as a function of grain size,
intensity and hardness of the incident spectrum.

\item We have developed a time-dependent radiative transfer code that
simultaneously follows and combines the process of dust destruction
through UV/ X-ray sublimation and through IFE with the evolution of
the ionization states of metals due to photoionization and their
relative radiative transitions. The code keeps track of the relative
fractions of metals in gaseous and in dust-depleted phase as a
function of space and time, and, as such, it follows the gradual
recycling into gas of metals while dust is being destroyed.

The code allows for any initial distribution of grain sizes and any
value for the dust-to-gas ratio and uses, through interpolation, the
tabulated opacities computed by Draine \& Lee (1983).  To reduce the
running times of the simulations, we have developed (and provided in
the Appendix) an approximate expression for the Planck averaged
absorption opacity that reproduces the numerical results by Draine \&
Lee within $\sim15\%$ in the range of grain sizes $0.01\le{}a_{-5}\le10^2$
and temperatures $10\le{}T\le5\times10^4$~K.

\item We have applied our code to show how GRBs affect the inferred
X-ray and optical extinction depending on the type of environment, and
how monitoring these quantities {\em during} the GRB event can provide
powerful clues on the characteristics of the environment itself. In
fact, we have shown how a measurement of X-ray absorption, $N_{\rm
H}$, and optical extinction, $A_V$, at a {\em single} time can yield
dust-to-gas ratios that are either lower or higher than the real
value, depending on the density and size of the absorbing region,
which determines the relative reduction in the high and low-energy
opacities.

Prompt and continuous time monitoring of the opacities in various
bands will be soon possible with {\em Swift} and REM. 

Whereas a knowledge of the type of environment in which GRBs
occur is relevant for constraining GRB progenitors, the inferred
metallicites and dust content are particularly important for a fair
and comprehensive study of high-redshift galaxies, whose inner and
denser cores are inaccessible to QSO absorption studies, and for which GRBs
constitute the only probe available so far.

\end{itemize}

\acknowledgements We are indebted with Bruce Draine for illuminating
discussions on the issue of dust destruction and with Fabrizio Fiore
for very useful discussions on observational issues related to dust.
We thank both of them, as well as an anonymous referee, for their careful
reading of our manuscript and insightful comments.  RP thanks the
Osservatorio Astronomico di Roma (Italy) and the California Institute
of Technology for their kind hospitality and financial support during
the time that this work was carried out.

\newpage

\centerline{APPENDIX}

\appendix

\section{Analytical approximation to the Planck averaged absorption
efficiency.}

We have approximated the Planck averaged absorption efficiencies
$\langle Q \rangle_T$ for graphite and silicates according to the
formula:

\begin{equation}
\langle{Q}\rangle_T = {{1}\over{1+D}}
\left[{{A\,a_{-5}}\over{1+A\,a_{-5}}}+D\left(
{{B\,a_{-5}}\over{1+B\,a_{-5}}}\right)^C\right]
\end{equation}
where
\begin{eqnarray}
{\rm Log}\,A &=& 
\left\{\begin{array}{ll}
A_0+A_1\xi & T<10 \\
A_2+A_3\xi+A_4\xi^2+A_5\xi^3+A_6\xi^4+A_7\xi^5 & 10\ge T\ge 10^5 \\
A_8+A_9\xi & T>10^5 \end{array}\right.\\
{\rm Log}\,B &=& 
\left\{\begin{array}{ll}
B_0+B_1\xi & T<10 \\
B_2+B_3\xi+B_4\xi^2+B_5\xi^3+B_6\xi^4+B_7\xi^5 & 10\ge T\ge 10^5 \\
B_8+B_9\xi & T>10^5 \end{array}\right. \\
\xi &=& {\rm Log}\,T \\
C &=& 300 \\
D &=& 6
\end{eqnarray}
and the coefficients $A_i$ and $B_i$ are given in Tab.~\ref{tab:a1}.

\begin{table}
\begin{center}
\begin{tabular}{c||rrrr}
      & Silicates ($A$) & Silicates ($B$) & Graphite ($A$) & Graphite ($B$)
\\ \hline \hline
$X_0$ & -4.85149        & -1.85901        & -4.79986       & -1.28807 \\
$X_1$ & 1.85976         & 1.55835         & 1.85669        & 1.18097  \\
$X_2$ & 6.53714         & 6.43232         & -6.44955       & -1.15262 \\
$X_3$ & -25.9463        & -19.4323        & 3.39539       & -0.370281 \\
$X_4$ & 24.7076        & 19.4393        & 1.28638       & 2.72991 \\
$X_5$ & -9.83150        & -8.11878        & -1.72185       & -1.76088 \\
$X_6$ & 1.73114         & 1.51996         & 0.529988       & 0.461037 \\
$X_7$ & -0.109282       & -0.104343       & -0.0507216     & -0.0419332 \\
$X_8$ & -16.1694        & -9.36205        & -4.58258       & -0.975760 \\
$X_9$ & 4.06422         & 2.90181         & 1.46046        & 1.09774
\end{tabular}
\end{center}
\caption{{Coefficients for the analytic approximation of the Planck 
averaged absorption efficiencies $\langle Q \rangle_T$ for graphite
and silicates. The first column gives the $A_i$ coefficients for
silicates, the second column gives the $B_i$ coefficients for
silicates, the third column gives the $A_i$ coefficients for graphite
and the fourth column gives the $B_i$ coefficients for graphite.}
\label{tab:a1}}
\end{table}

\newpage

\begin{figure}
\centerline{\epsfysize=5.7in\epsffile{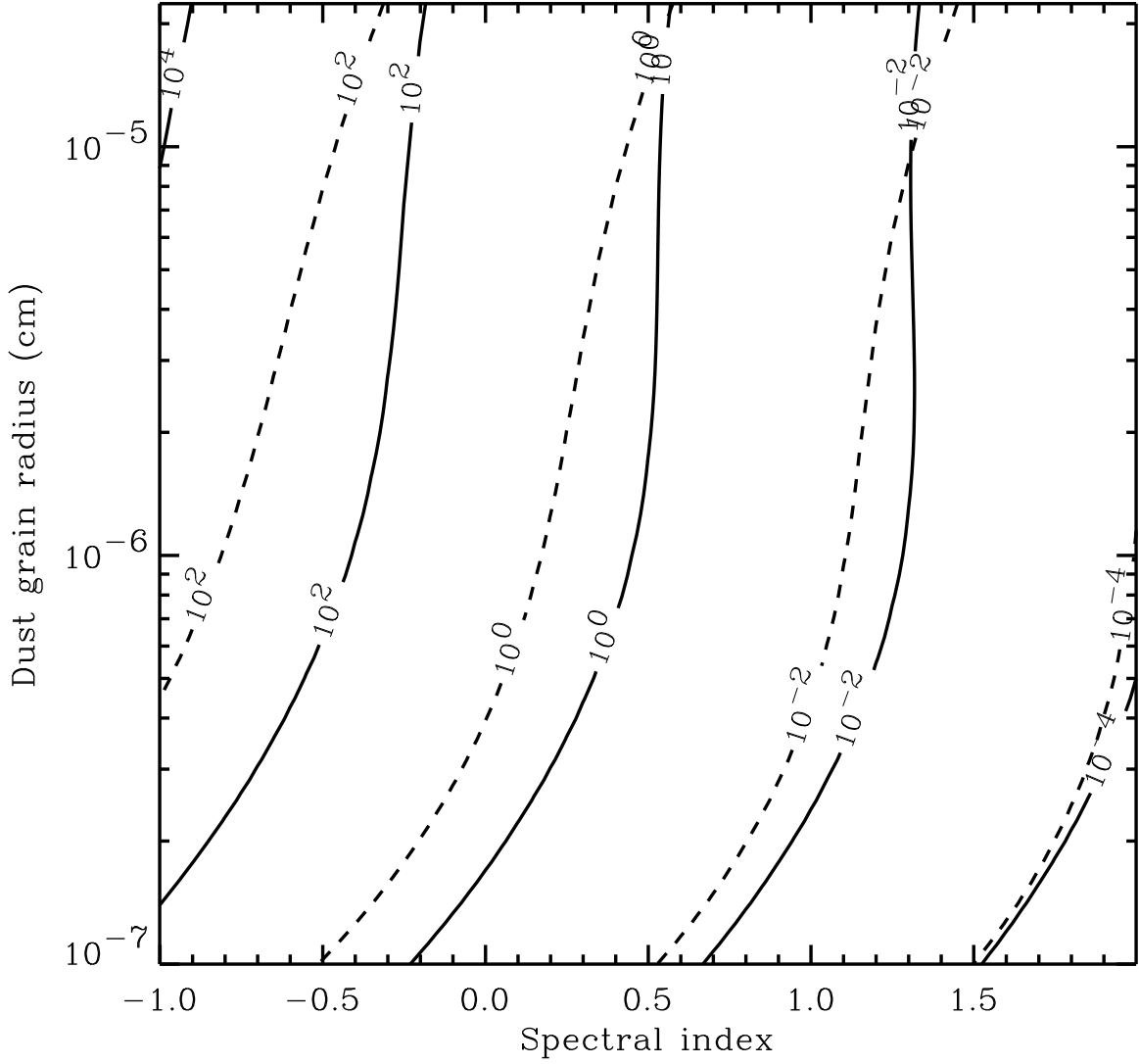}}
\caption{Efficiency of the X-ray heating process with respect
to the UV one.  The figure  shows contour plots of the ratio of Eq. (9) for
different spectra and grain sizes and for both types of grain
composition, i.e. graphite (solid lines) and silicates (dashed lines).}
\label{fig:eff}
\end{figure}

\begin{figure}
 \centerline{\epsfysize=5.7in\epsffile{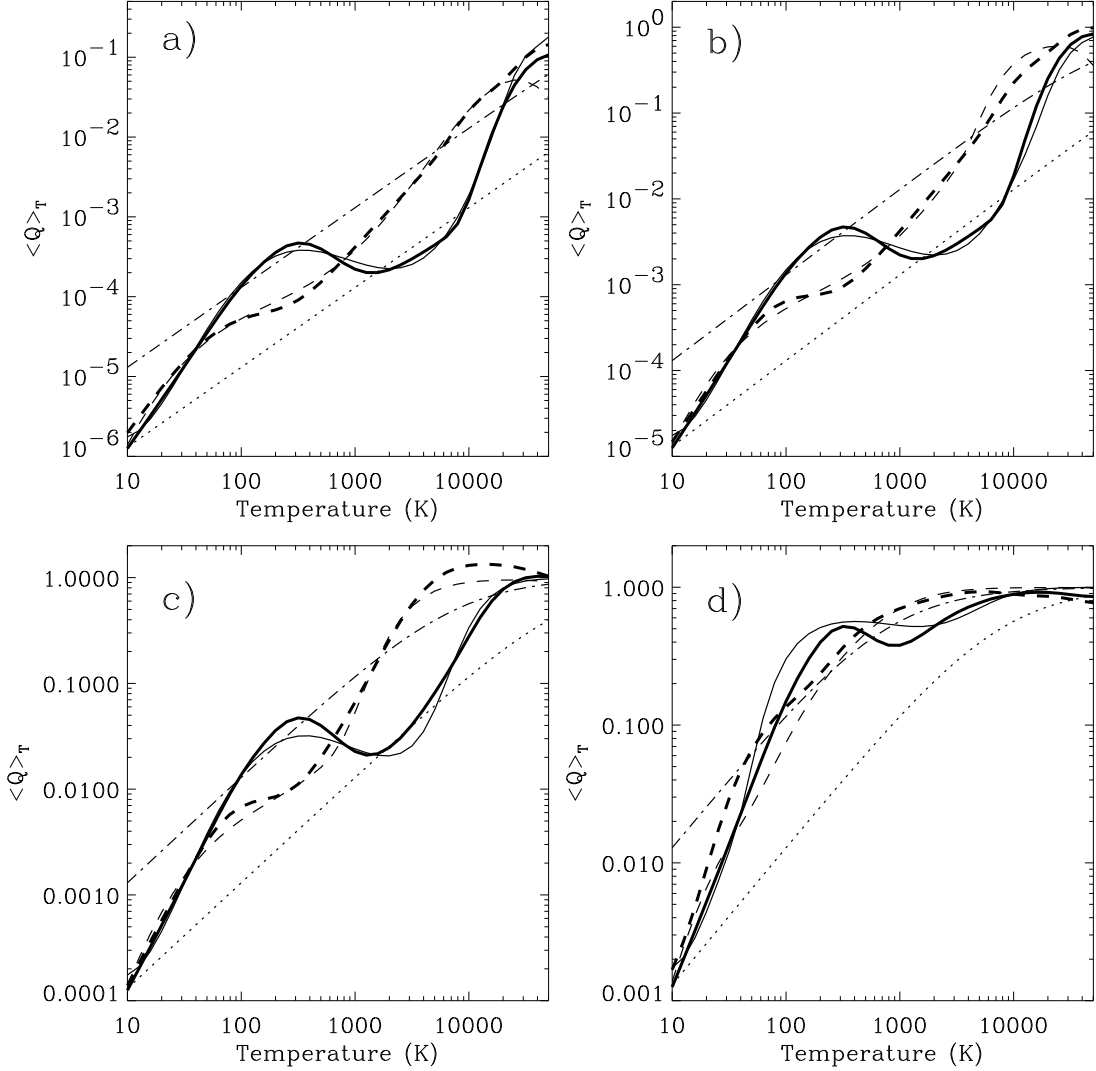}}
\caption{The thick lines show the Planck-averaged absorption
efficiency $<Q_T>$ from the numerical models of Draine \& Lee (1984),
the solid lines for silicates and the dashed line for graphite and for
different values of the grain size: $a_{-5}=0.01$ (panel a), $a_{-5}=0.1$
(panel b), $a_{-5}=1$ (panel c), $a_{-5}=10$ (panel d). The approximation to
$<Q_T>$ by Waxman \& Draine (2000) is shown with the dotted lines for
silicates and dotted-dashed lines for grafite in each of the panels
for the same values of the grain sizes. Finally, the thin lines show
the approximation to $<Q_T>$ that we provided in Appendix A, the solid
lines for silicates and the dashed lines for graphite, and again for
the same values of the grain sizes.}
\label{fig:qt}
\end{figure}

\begin{figure}
 \centerline{\epsfysize=5.7in\epsffile{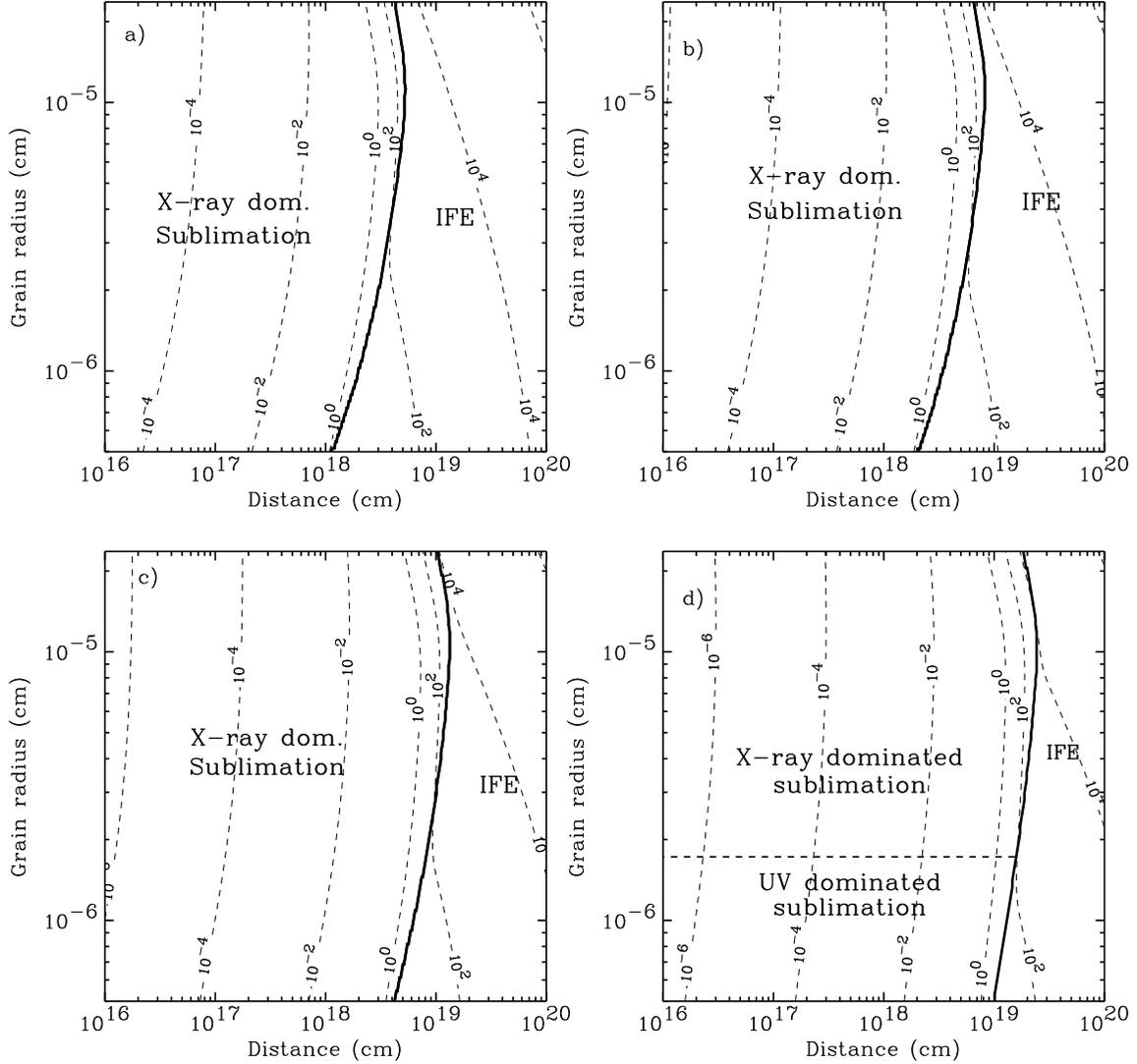}}
\caption{The relative importance of the three processes
of grain sublimation (UV heating, X-ray heating and IFE) as a function
of the size of the grain and the intensity of the flux (or
equivalently the distance of the grain from the source) for silicates
dust particles (MgFeSiO4) and a power-law incident spectrum
$F(\nu)\propto\nu^{-\alpha}$.  The panels show the regions in the
parameter space where each mechanism is dominant over the remaining
two.  Different panels show different spectral shapes: panel (a) has
$\alpha=-0.25$, in panel (b) $\alpha=0$, in panel (c) $\alpha=0.25$,
and finally panel (d) has been computed with $\alpha=0.5$. The softer
the spectrum, the more important UV sublimation becomes with respect
to the other two processes. The dashed labeled contours show the grain
destruction timescale in seconds, for a 1~eV---100~keV luminosity
$L=10^{50}$~erg~s$^{-1}$, typical for the prompt phase of GRBs.}
\label{fig:sil}
\end{figure}

\begin{figure}
 \centerline{\epsfysize=5.7in\epsffile{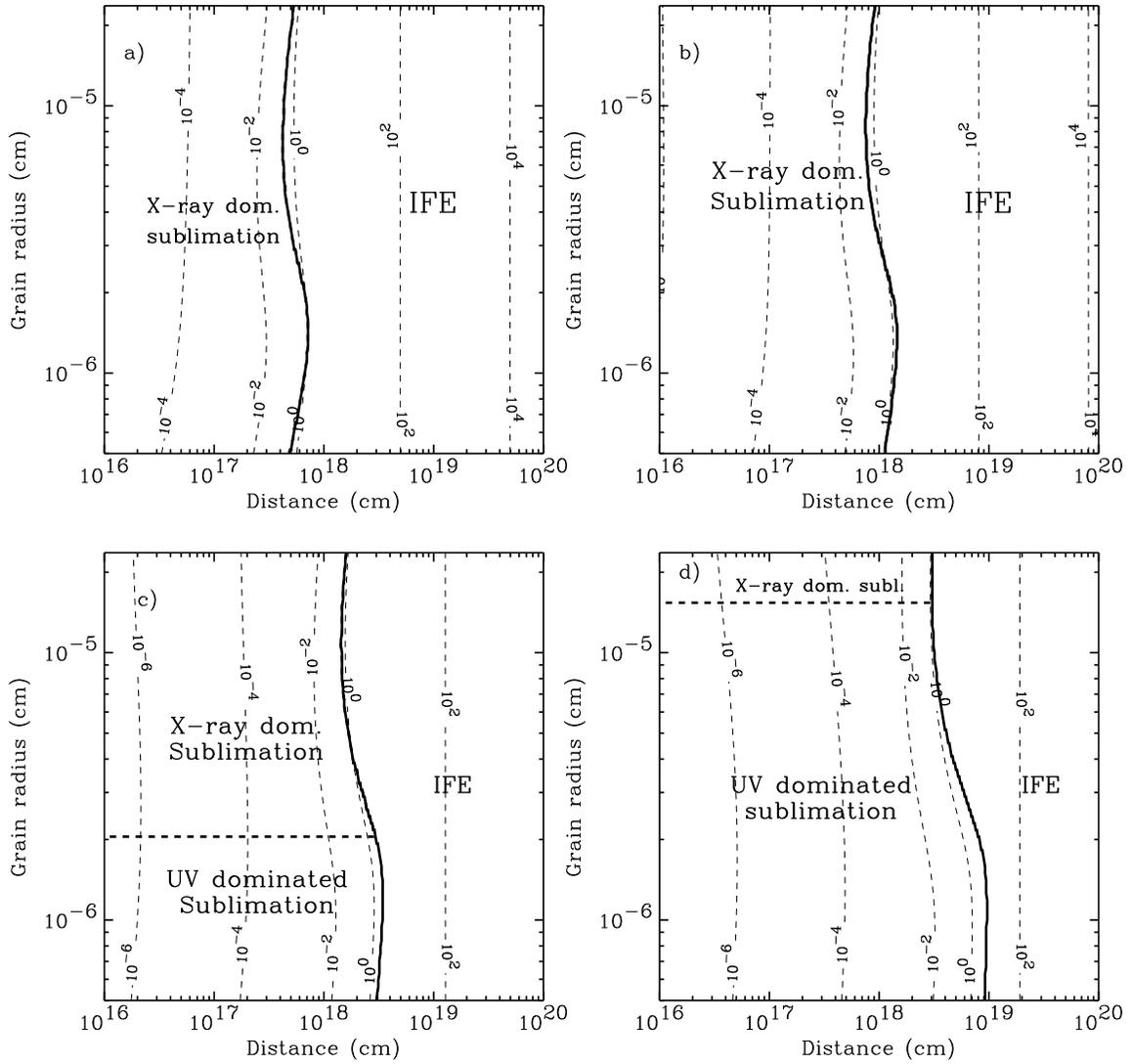}}
\caption{Same as in Fig. 3 but for graphite.}
\label{fig:gra}
\end{figure}

\begin{figure}
 \centerline{\epsfysize=5.7in\epsffile{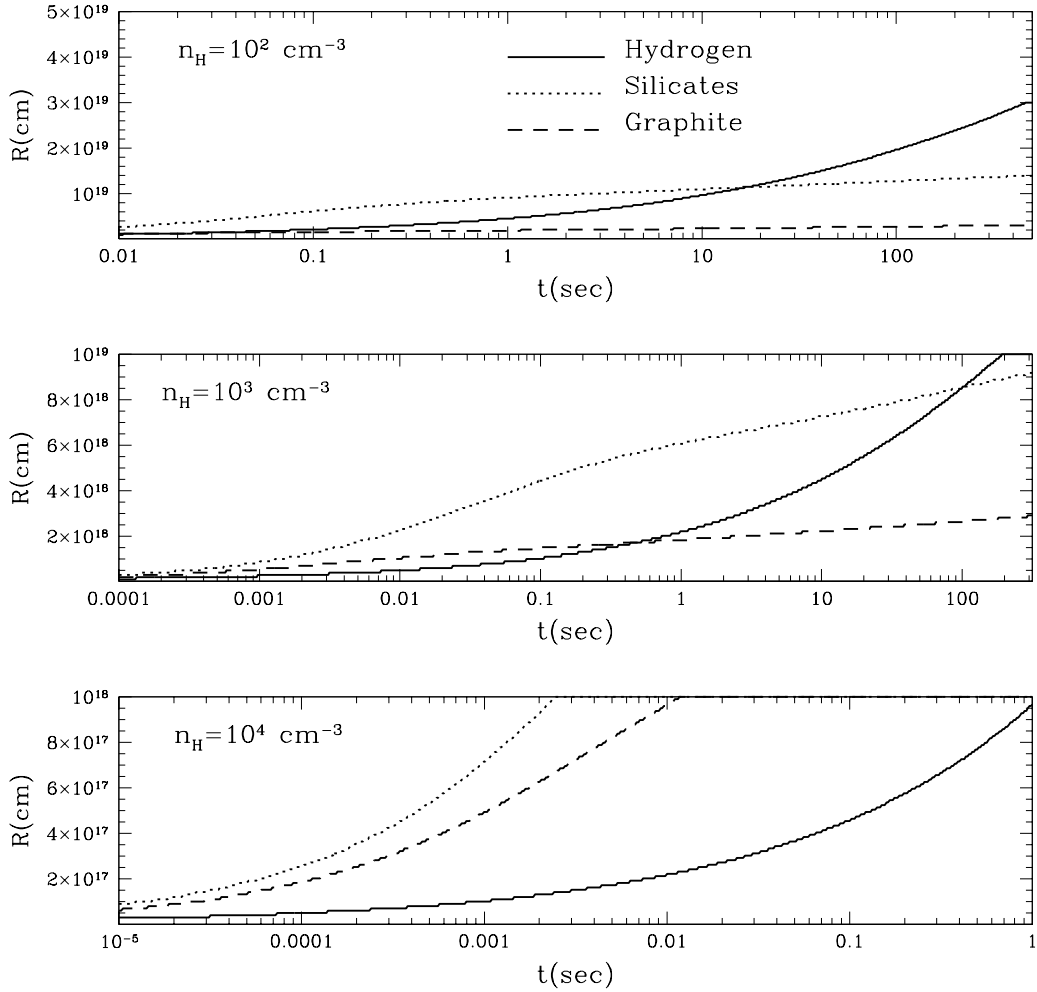}}
\caption{Time evolution of the destruction fronts in environments
characterized by different densities.  The ionization front of
Hydrogen is defined by the condition that the fractional abundance of
neutral Hydrogen falls below $10^{-5}$, while the destruction fronts
of silicates and graphite are defined by the condition of complete
sublimation of the largest grains at that radius. In this simulation,
the incident flux is a power law with $\alpha=0.5$ and normalization
as in Fig.~2.}
\label{fig:df}
\end{figure}

\begin{figure}
 \centerline{\epsfysize=5.7in\epsffile{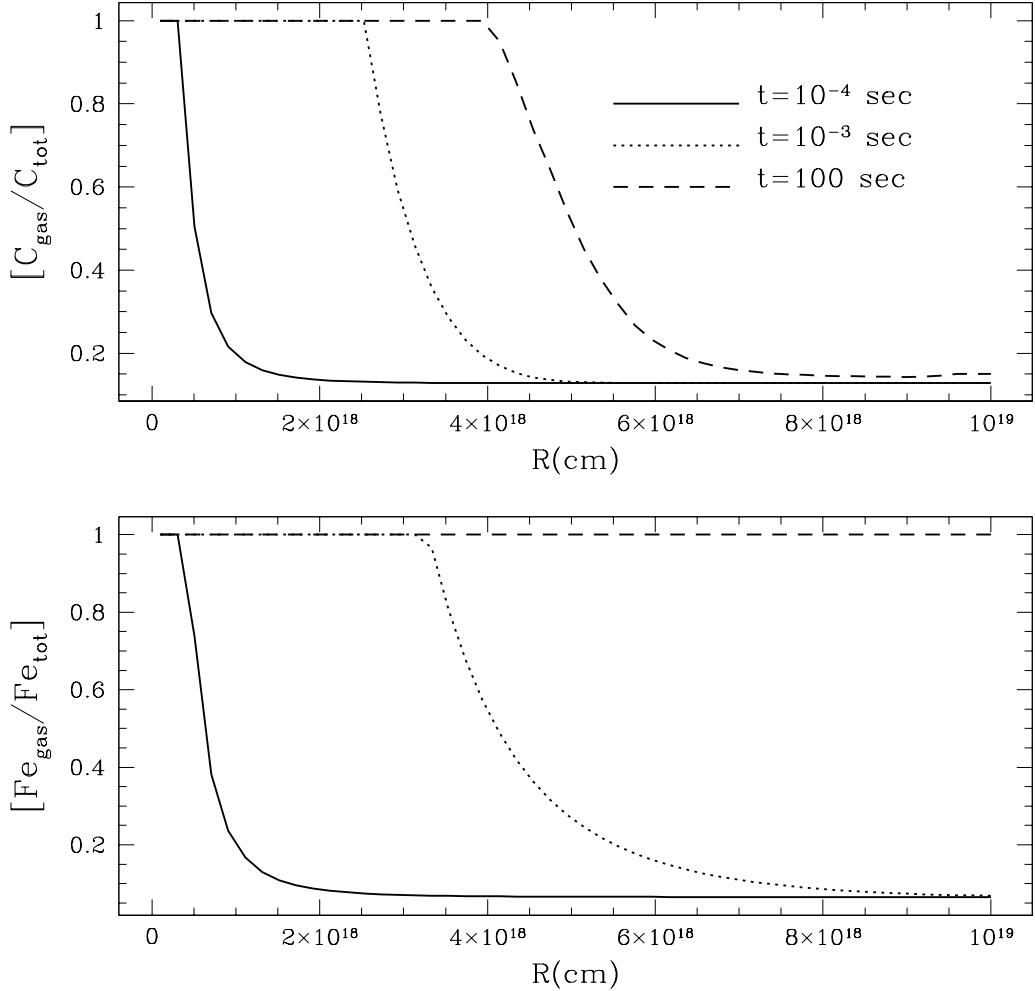}}
\caption{The gradual recycling of metals from dust into gas during the passage
of the radiation flux is illustrated here. The two panels show the
fractional abundance in gaseous phase of Carbon (released from the
destruction of graphite) and Iron (released from the destruction of
silicates) at various times within the region surrounding the
source. These results have been obtained from a simulation in a region
with density $n_{\rm H}=10^3$ cm$^{-3}$ and radius $R=10^{19}$ cm.}
\label{fig:abun}
\end{figure}

\begin{figure}
 \centerline{\epsfysize=5.7in\epsffile{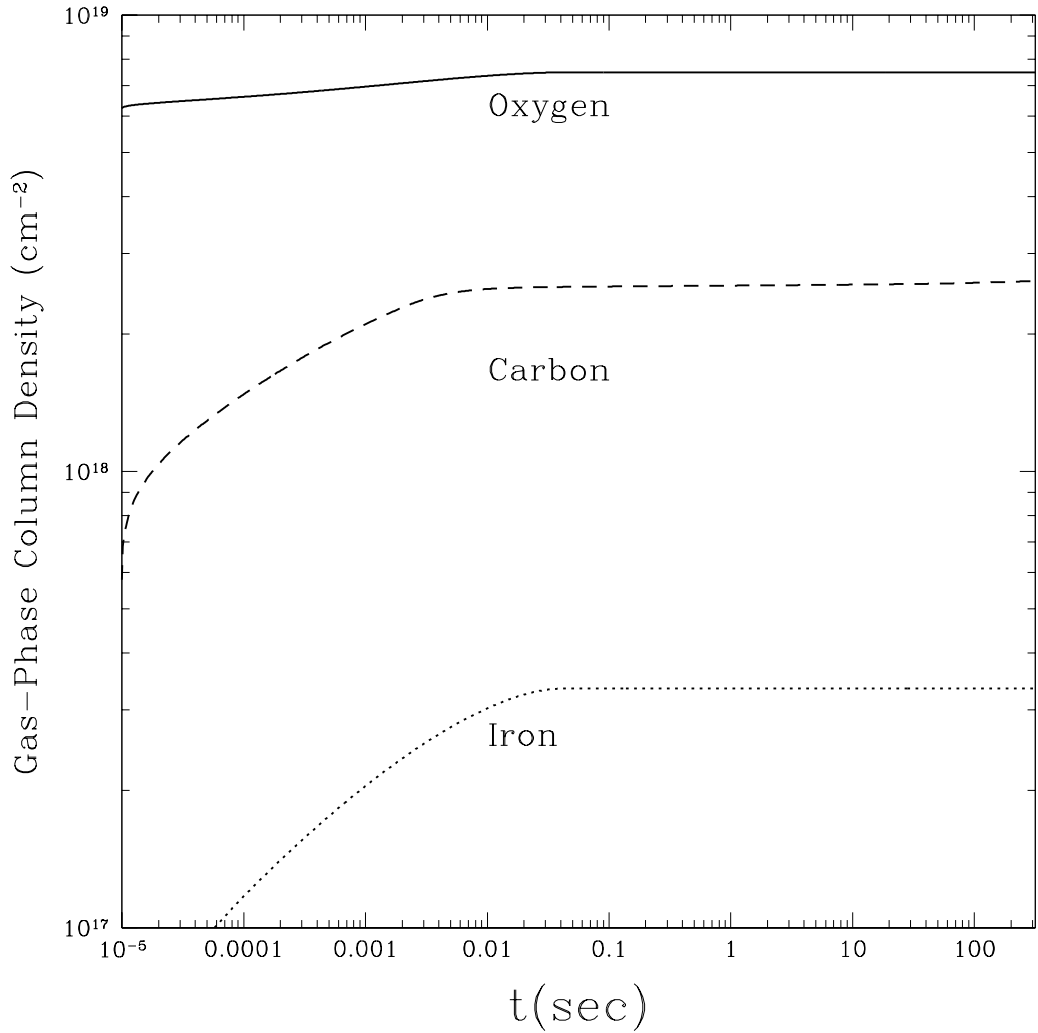}}
\caption{Column densities of Iron, Oxygen and Carbon in gaseous phase 
as the radiation front propagates destroying dust and recycling the
depleted metals into gas. Density and radius of the region are the
same as in Fig.~6. }
\label{fig:col}
\end{figure}

\begin{figure}
 \centerline{\epsfysize=5.7in\epsffile{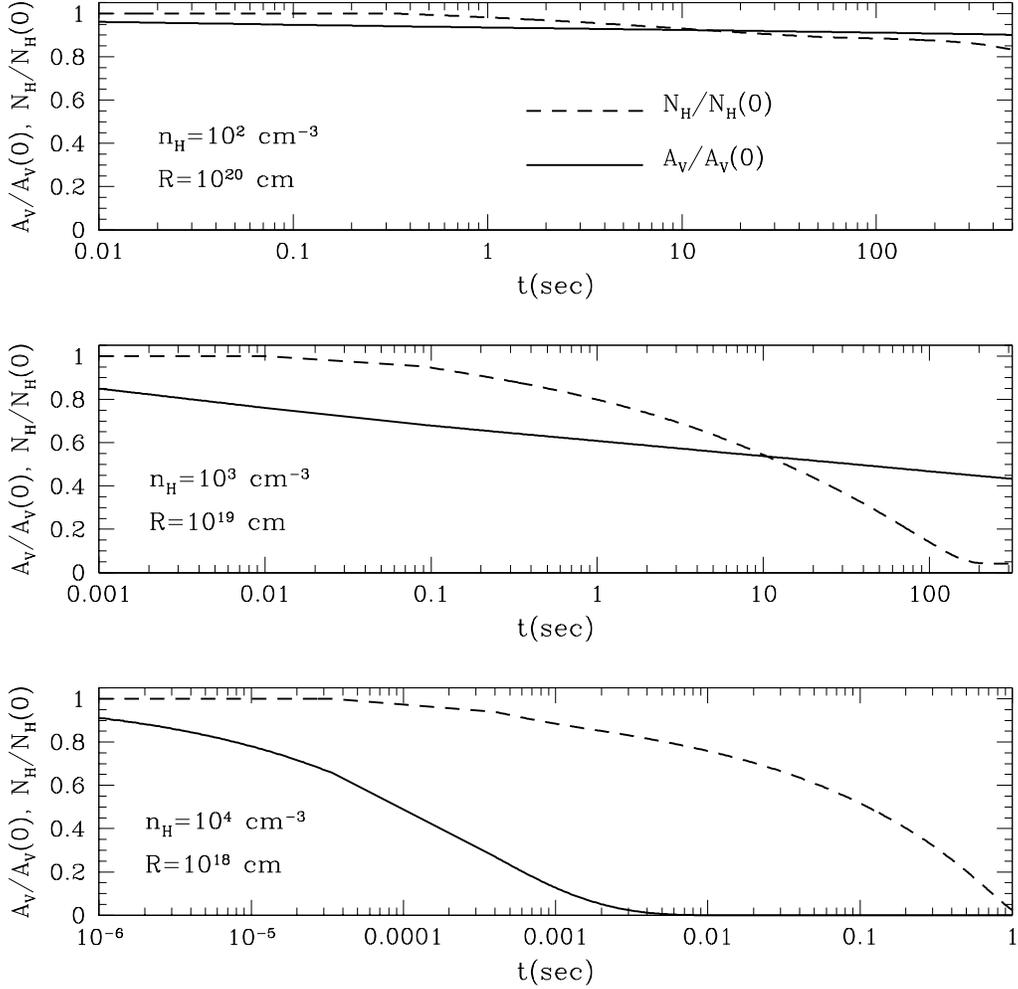}}
\caption{Time evolution of the X-ray and optical extinction in 
different types of environments. In all cases the initial Hydrogen
column density is $N_{\rm H}(0)=10^{22}$ cm$^{-2}$ while the initial
optical extinction is $A_V(0)=4.5$ mag. The time evolution of $N_{\rm
H}(t)$ and $A_V(t)$ is very sensitive to the characteristics (density,
size) of the region surrounding the source. In this simulation,
the incident flux is a power law with $\alpha=0.5$ and normalization
as in Fig. 2.}
\label{fig:nhav}
\end{figure}

\begin{figure}
 \centerline{\epsfysize=5.7in\epsffile{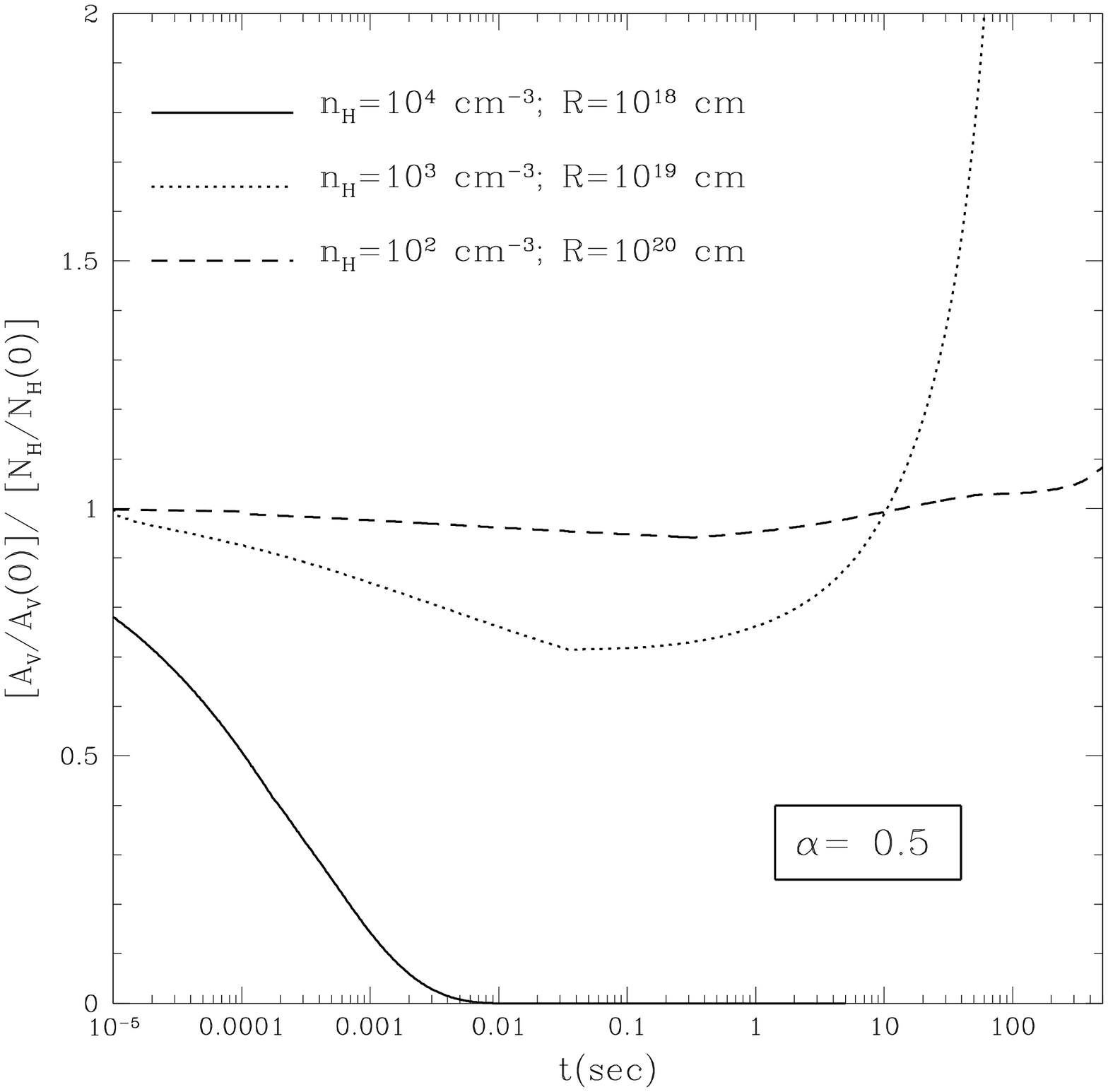}}
\caption{The ratio between the time-variable optical and 
X-ray extinction is shown here for the three types of environments
and illuminating flux considered in Fig.8.}
\label{fig:nhavr}
\end{figure}

\begin{figure}
 \centerline{\epsfysize=5.7in\epsffile{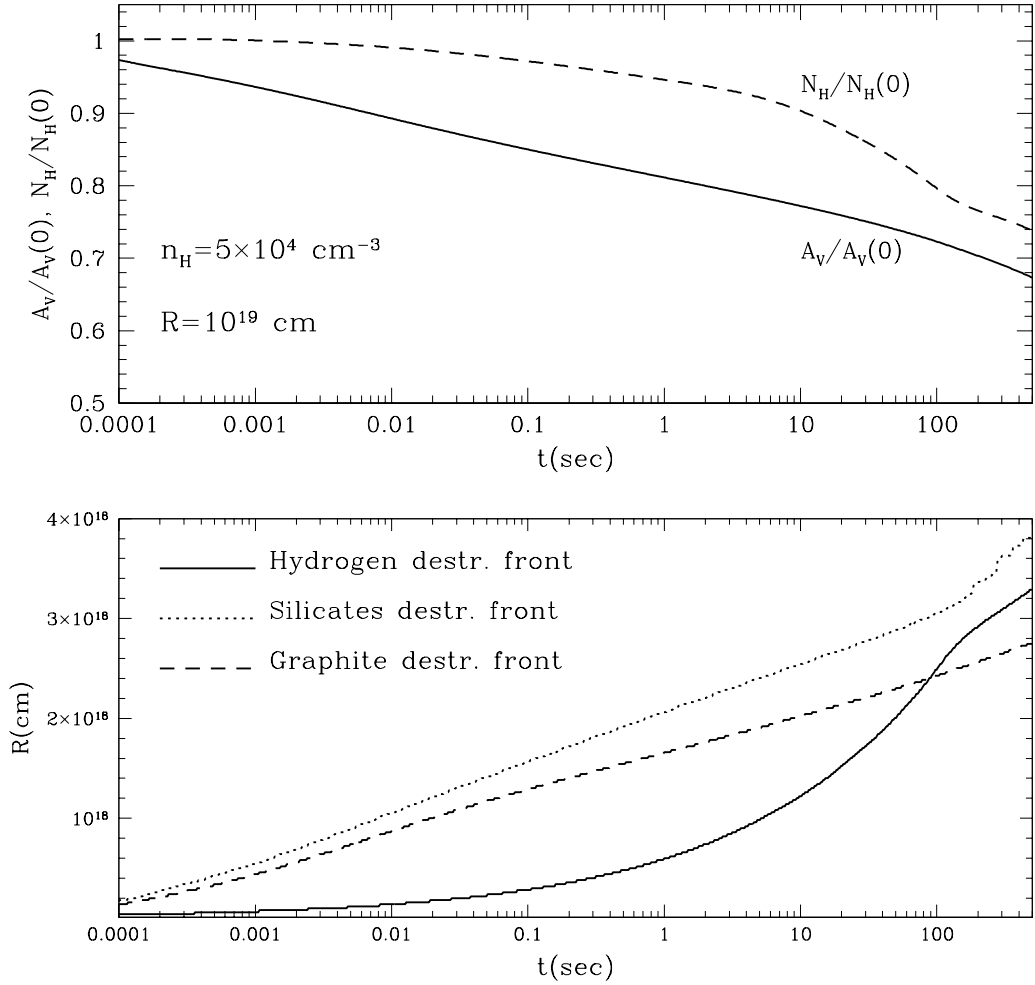}}
\caption{Destruction fronts of Hydrogen and dust (top panel) and time
evolution of the X-ray and the optical extinction for a rather large
and very dense region. The size of the region is the same as in the
middle panel of Figs. 5 and 8 (and the same is
also the incident spectrum), but the density is 50 times higher. This
results in an enhancement of the separation between the dust
destruction front and the ionization front that lags behind it at high
densities.}
\label{fig:dfnh}
\end{figure}

\begin{figure}
 \centerline{\epsfysize=5.7in\epsffile{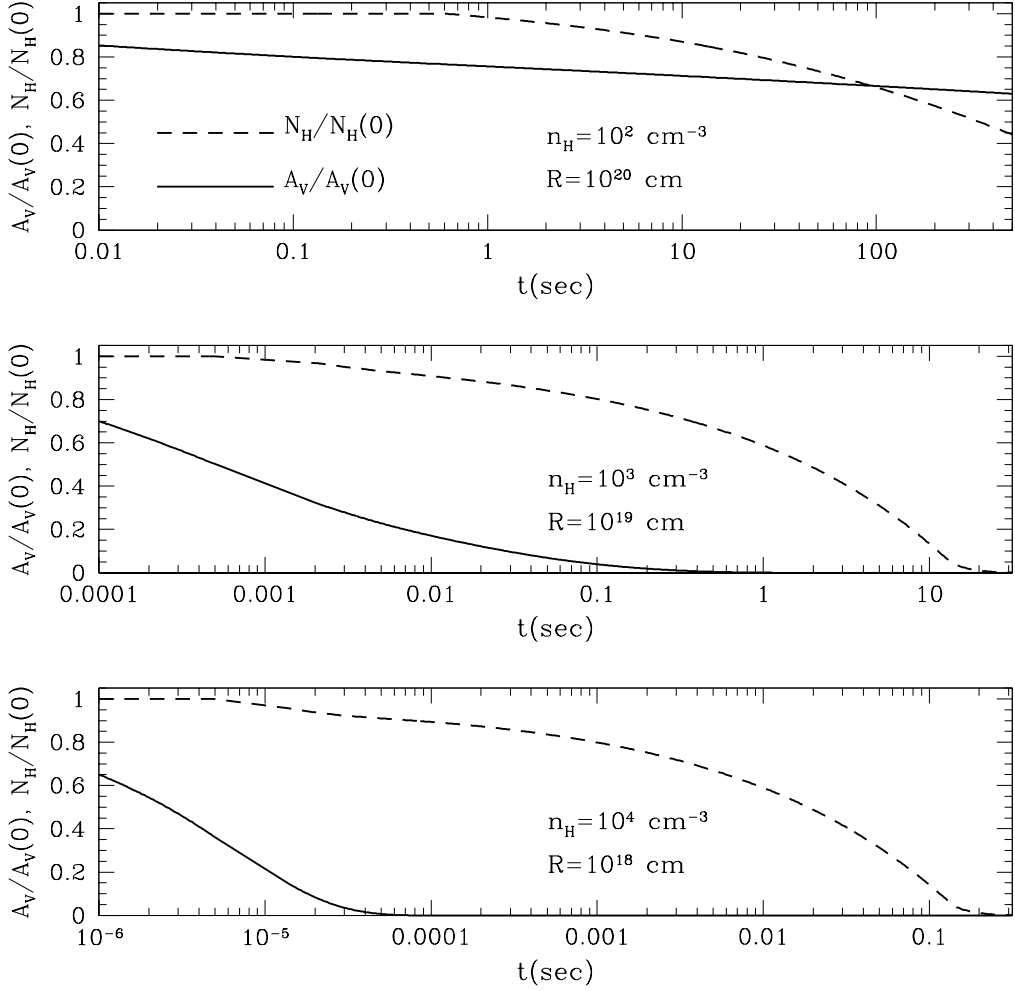}}
\caption{Time evolution of the X-ray and optical extinction in 
the same types of environments considered in Fig.8  but for a softer
illuminating spectrum, with spectral index $\alpha=1$. }
\label{fig:nhaval1}
\end{figure}

\begin{figure}
 \centerline{\epsfysize=5.7in\epsffile{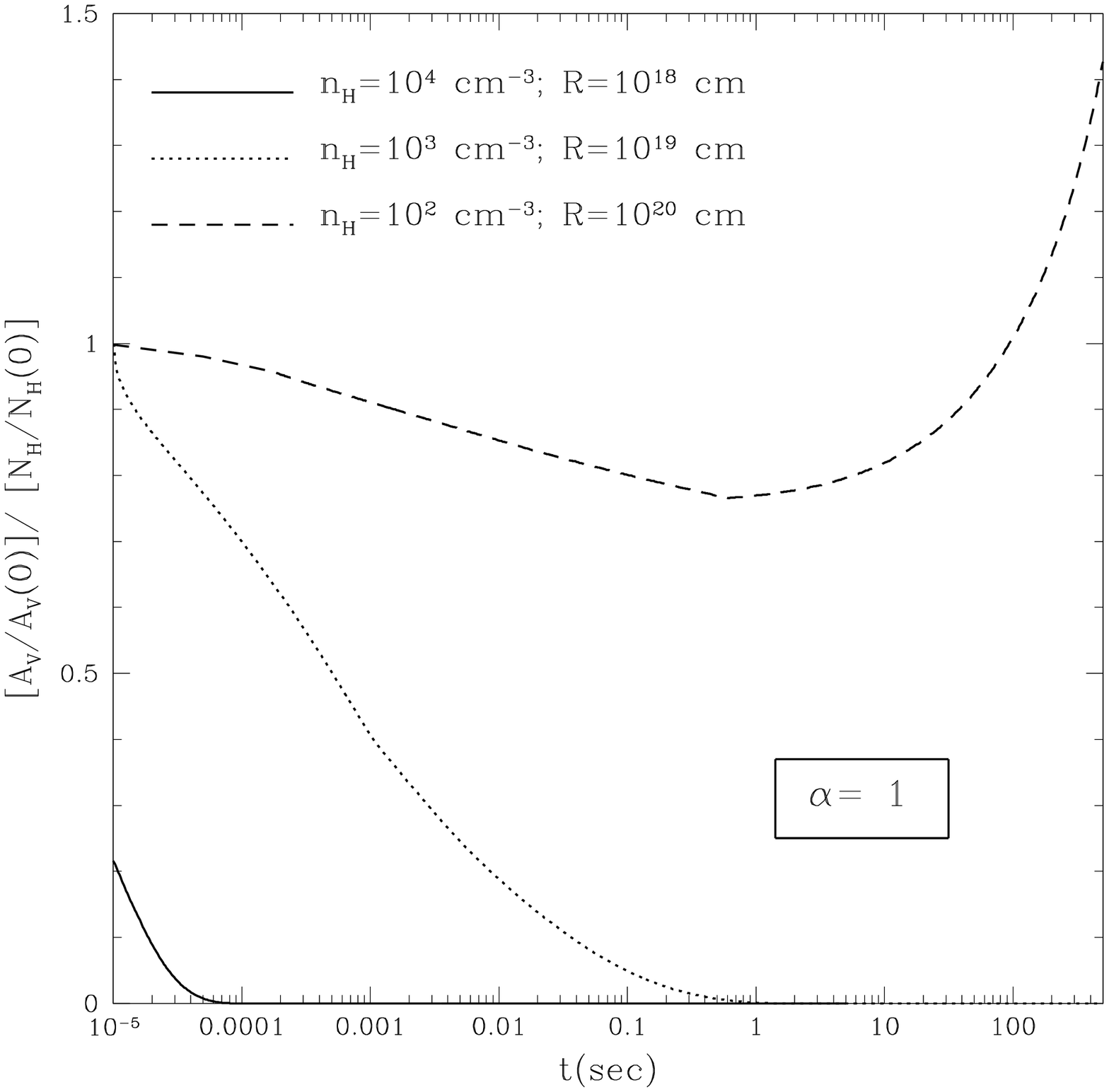}}
\caption{The ratio between the time-variable optical and 
X-ray extinction is shown here for the three types of environments
and illuminating flux considered in Fig.11.}
\label{fig:nhavr2}
\end{figure}

\end{document}